%% file: ms.tex
\documentclass[12pt,preprint]{aastex}
\usepackage{psfig,amsmath}

\shorttitle{E0102 overview paper}
\shortauthors{Flanagan, et al.}

\begin{document}

\title{{\it Chandra} High-Resolution X-Ray Spectrum of Supernova Remnant
1E\,0102.2-7219}

\author{K.A.~Flanagan, C.R.~Canizares, D.~Dewey, J.C.~Houck, 
  A.C.~Fredericks, M.L.~Schattenburg, T.H.~Markert, 
D.S.~Davis\footnote{Current address: Laboratory for 
High Energy Astrophysics, Code 661, NASA/Goddard Space Flight Center, 
Greenbelt, MD 20771}}
\affil{Center for Space Research, Massachusetts Institute of
Technology, Cambridge, MA 02139}

\email{kaf@space.mit.edu}

\begin{abstract}

{\it Chandra} High Energy Transmission Grating Spectrometer
observations of the supernova remnant 1E\,0102.2-7219 in the Small
Magellanic Cloud reveal a spectrum dominated by X-ray emission lines
from hydrogen-like and helium-like ions of oxygen, neon, magnesium and
silicon, with little iron. The dispersed spectrum shows a series of
monochromatic images of the source in the light of individual spectral
lines.  Detailed examination of these dispersed images reveals Doppler
shifts within the supernova remnant, indicating bulk matter velocities
on the order of 1000~km~s${}^{-1}$. These bulk velocities suggest an
expanding ring-like structure with additional substructure, inclined
to the line of sight. A two-dimensional spatial/velocity map of the
SNR shows a striking spatial separation of redshifted and
blueshifted regions, and indicates a need for further investigation
before an adequate 3D model can be found.  The radii of the ring-like
images of the dispersed spectrum vary with ionization stage,
supporting an interpretation of progressive ionization due to passage
of the reverse shock through the ejecta.  Plasma diagnostics with
individual emission lines of oxygen are consistent with an ionizing
plasma in the low density limit, and provide temperature and
ionization constraints on the plasma.  Assuming a pure metal plasma,
the mass of oxygen is estimated at $\sim$6~M$_{\odot}$, consistent
with a massive progenitor.

\end{abstract}
\keywords{ISM: individual (1E 0102.2--7219) --- Magellanic Clouds --- plasmas --- supernova remnants --- techniques: spectroscopic --- X-rays: ISM}


\section{Introduction}

	The SNR 1E\,0102.2-7219 (E0102) is a well studied member of
the oxygen rich class of supernova remnants located in the Small
Magellanic Cloud (SMC). \citet{Gaetz00} reported spectrally resolved
imaging from {\it Chandra's} ACIS detector, which shows an almost
classic, text-book SNR with a hotter outer ring identified with the
forward shock surrounding a cooler, denser inner ring which is
presumably the reverse-shocked stellar ejecta.  \citet{Hughes00}
combined the {\it Chandra} image with earlier {\it Einstein} and {\it
ROSAT} images to measure X-ray proper motions, which give an expansion
age of $\sim$1000 yr, consistent with earlier estimates based on
optical measurements of oxygen-rich material~\citep{Tuohy83, Dopita81,
Hayashi94, Gaetz00}.  By contrast, \citet{Eriksen01} estimate a free
expansion age of 2100~years.  \citet{Hughes00} deduce that a
significant fraction of the shock energy has gone into cosmic
rays. Composite X-ray spectra of moderate resolution have been
obtained for the whole remnant with ASCA~\citep{Hayashi94} and
XMM--Newton~\citep{Sasaki01}. These observations confirm that a single
component non-equilibrium plasma is inadequate to account for the
global SNR spectrum.  A high resolution X-ray spectrum obtained with
the reflection grating spectrometer of XMM--Newton reveals a wealth of
individual lines of C, O, Ne, Mg and Si~\citep{Rasmussen01}.  Line
ratios from these spectra confirm non-equilibrium ionization (NEI)
conditions consistent with an ionizing plasma in the low density
limit.

We report on the high resolution X-ray spectrum of 1E\,0102.2-7219
obtained with the {\it Chandra} High Energy Transmission Grating
Spectrometer; a portion of the spectrum is shown in
Figure~\ref{disperse} (preliminary reports were presented
in~\citet{Flanagan01, Canizares01, Flanagan03}).  Individual X-ray lines of oxygen, neon,
magnesium and silicon echo the sharply defined ring structure seen in
direct {\it Chandra} images.  Notably weak are lines of highly ionized
iron. This lack of strong iron lines is fortuitous, since the X-ray
spectra of SNRs are commonly dominated (and complicated) by an ``iron
forest'' of lines between 7~{\AA} and 18~{\AA}.  The spatial and
spectral signatures of E0102 make it a unique candidate for the HETG
-- bright, sharp spatial features arrayed in a narrow ring around a
sparsely filled interior, combined with a spectrum dominated by a
relatively small number of discrete lines.  As a result, the HETG
dispersed line images have only a small amount of overlap between
them, enabling straightforward analysis of the individual lines.

 
The analysis we present follows four broad areas: spectral lines and
line fluxes, ionization structure, Doppler shifts and
three-dimensional modeling.  Section~\ref{Observations} describes the
observations and data processing. Section~\ref{Spectrum} introduces
the spectrum, identifies line series, presents line images
(Section~\ref{Images}), and discusses flux measurements
(Section~\ref{Flux}) which can be cleanly obtained through the
relative simplicity of the dispersed spectrum.  We follow this with a
discussion of plasma diagnostics (Section~\ref{Plasma}) and an
estimate of oxygen ejecta mass and associated progenitor models
(Section~\ref{Ejecta_mass}).  Section~\ref{Shock} addresses the
ionization structure of the SNR, and relates it to passage of the
reverse shock through the ejecta.  Section~\ref{Doppler} outlines the
analysis of Doppler shifts, beginning with techniques and measurements
on specific sectors of the SNR, and branching out
(Section~\ref{Dewey}) to a Doppler velocity imaging technique which
yields a ``Doppler map'' of the entire SNR.  Section~\ref{Model}
introduces a simple three-dimensional model as a first attempt to
account for the spatial and velocity structure of the SNR.
Sections~\ref{Discussion} and~\ref{Summary} discuss and summarize the
findings.

\section{Observations and Data Analysis} \label{Observations}

         The supernova remnant E0102 was observed with the {\it
Chandra} High Energy Transmission Grating Spectrometer (HETGS)
\citep{Canizares00, Canizares03} in two observation intervals as part
of the guaranteed time observation program. Details of the observation
are given in Table~\ref{tbl-obs}.  The instrument configuration
included HETG with the Advanced CCD Imaging Spectrometer
(ACIS-S)~\citep{Garmire03, Burke97}. These two observations of
1E\,0102.2-7219 had slightly different roll angles and aim points and
were independently treated in our analysis.

\placetable{table1}

        The data were processed using standard CXC pipeline software
(version R4CU5UPD8.2), employing calibration files available in
February, 2001.  Processing of the data included running
acis\_process\_events to correctly assign ACIS pulse height to the
events (needed for proper order sorting), and filtering the data for
energy, status and grade (0,2,3,4,6).  Since the supernova remnant is
extended, a customized region mask was created to ensure that all
source photons were captured, both in the zeroth order (undispersed)
image and along the dispersion axes.  Further processing included
aspect correction, selection of good time intervals, removal of
detector artifacts (hot pixels and streaks), and selection of first
order photons.  At the end of processing, the net live time was
135.5~ks, the bulk of it (86.9~ks) from the first observation
interval, Obsid 120.

        The HETG consists of two independent sets of gratings with
dispersion axes oriented at angles that differ by $\sim$10$^\circ$.
The medium energy gratings (MEG) cover an energy range of 0.4~--~5~keV
and have half the dispersion of the high energy gratings (HEG), which
provide simultaneous coverage of the range 0.9~--~10~keV.  The
gratings form an undispersed image at the pointing position (the
zeroth order image) with {\it Chandra's} full spatial resolution, and
with spectral information limited to the moderate resolution provided
by the ACIS detector. The dispersed photons provide the high
resolution spectrum.  The different dispersion directions and two
dispersion wavelength scales provide redundancy as well as a means of
resolving spectral/spatial confusion problems associated with extended
sources such as supernova remnants (discussed in more detail in
Section~\ref{Doppler}). The high resolution spectra from +1 and $-1$
orders are discriminated from overlapping higher orders by using the
moderate energy resolution of the ACIS-S.  Further details of the
instrument can be found in \citet{Markert94}, \citet{POG02} and at
{\tt http://space.mit.edu/HETG}.

	After eliminating higher orders, approximately 47\% of the
detected photons in the spectrum were in zeroth order, 40\% in MEG $\pm$1
orders, and 13\% in HEG $\pm$1 orders. The MEG $-1$ order had a higher
count rate than the +1 order, attributable to the presence of a
backside illuminated CCD, which has much higher detection efficiency
for the bright, low energy oxygen lines than the frontside illuminated
CCD used in the +1 order.  The approximate breakdown of dispersed and
undispersed photons in the spectrum is shown in Table~\ref{tbl-obs}.

\section{The High Resolution X-ray Spectrum} \label{Spectrum}

\begin{figure*}
\figurenum{1}
\psfig{file=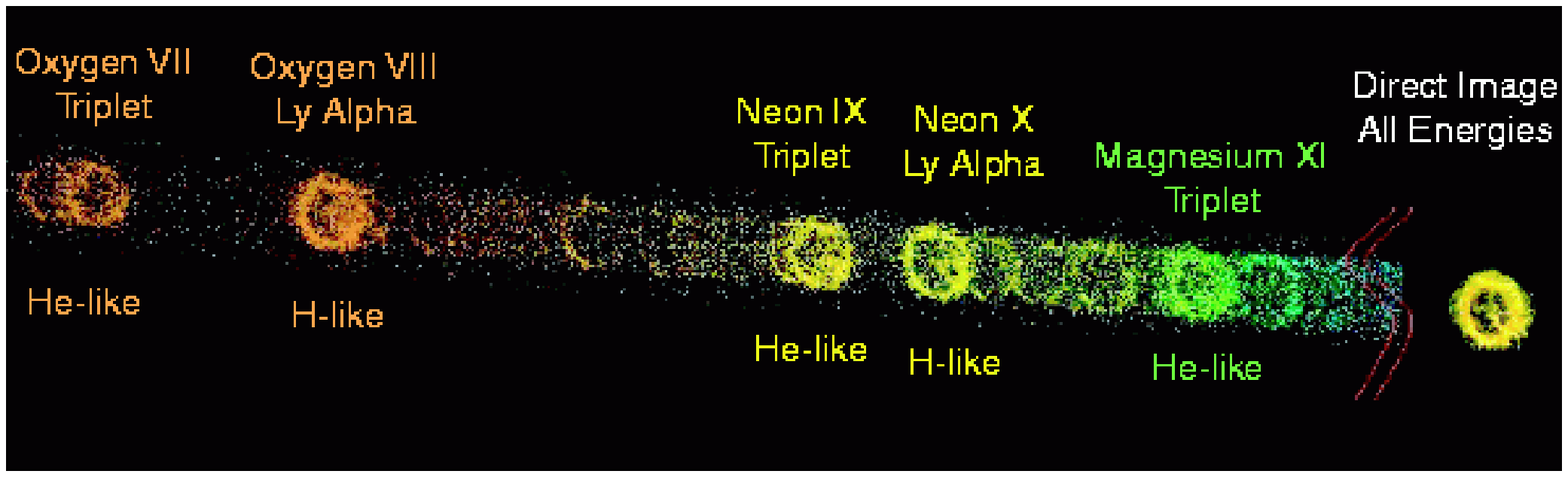,width=1.00\textwidth}
\figcaption[f1.eps]
{Dispersed high resolution spectrum of 1E\,0102-7219.  Shown here is 
a portion of the Medium Energy Grating (MEG)~-1~order, color coded 
to suggest the ACIS energy resolution.  At right in the figure 
(with different intensity scaling) is the zeroth order, which combines 
all energies in an undispersed image.  Images formed in the light of 
strong X-ray emission lines are labeled.
The dispersed spectrum is truncated; the resonance line of Si\,XIII is
not shown in the figure.
\label{disperse}}
\end{figure*}

        Figure~\ref{disperse} shows a portion of the high resolution
spectrum from the $-1$ order of the MEG gratings, taken during the
first observation interval, Obsid~120.  The dispersed spectrum is
analogous to a spectro-heliogram, showing a series of monochromatic
images of the source in the light of individual spectral lines. The
HETGS spectrum is dominated by lines of highly ionized oxygen, neon
and magnesium from ionization stages in which only one (hydrogen-like)
or two (helium-like) electrons remain. These are states which are
long-lived under conditions typical of supernova remnants. Also
present is a helium-like line of silicon (truncated from
Figure~\ref{disperse}), but notably weak are lines of highly ionized
iron (i.e., Fe~XVII and Fe~XVIII). Relative to the strong lines in the
spectrum, the continuum component is weak.

\begin{figure*}
\figurenum{2}
\psfig{file=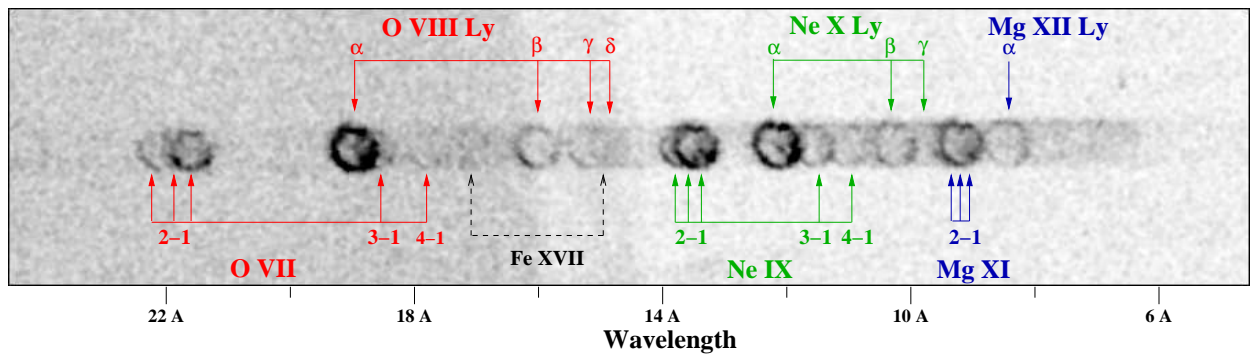,width=1.00\textwidth}
\figcaption[f2.eps]{Dispersed high resolution spectrum of
1E\,0102-7219 from MEG $-1$ order, emphasizing faint X-ray lines. The
image is summed from the two observation intervals, Obsid~120 and
Obsid~968. Transitions from upper levels for various elements and
ionization species are indicated. The expected locations of normally
bright Fe~XVII lines are indicated; Fe is obviously weak in the
spectrum. Detector artifacts, such as the chip gaps appearing near
O\,VIII~Lyman~$\beta$ and midway between O\,VIII~Lyman~$\delta$ and
Ne\,IX~forbidden, have not been removed. Silicon and faint magnesium 
transitions are not marked. \label{transitions}}
\end{figure*}

	The spectrum reveals multiple lines from the various ions.
Several of these transitions are indicated in
Figure~\ref{transitions}.  The hydrogen-like oxygen lines indicate
transitions from upper levels n~=~2,~3,~4,~5 to n~=~1 (i.e.,
O\,VIII~Lyman~$\alpha$, $\beta$, $\gamma$, and $\delta$,
respectively).  The hydrogen-like neon lines include n~=~2,~3,~4 to
n~=~1 transitions (Ne\,X~Lyman~$\alpha$, $\beta$, and $\gamma$).  The
helium-like oxygen (O\,VII) and neon (Ne\,IX) lines include the
n~=~2,~3,~4 to n~=~1 transitions. In each case, the triplet of lines
from n~=~2 to n~=~1 is characterized by bright forbidden (1$s^2$
$^1S_0$~--~1s2s $^3S_1$) and resonance lines (1$s^2$
$^1$S$_0$~--~1s2p~$^1P_1$) and a weak intercombination line (1$s^2$
$^1S_0$~--~1s2p~$^3P_{2,1}$). For the hydrogen-like magnesium ion
(Mg\,XII), only n~=~2,~3 to n~=~1 transitions are detected, and for
the helium-like magnesium (Mg\,XI), n~=~2,~3,~4 to n~=~1 transitions
are detected. Finally, the helium-like Si\,XIII transition from n~=~2
to n~=~1 is detected.  Note that the silicon transition and fainter
magnesium transitions are not marked in Figure~\ref{transitions}.


	Lines of L-shell transitions of Fe, often quite strong in SNR
spectra, are weak in the E0102 spectrum. (Locations where bright Fe
lines would typically appear are depicted in Figure~\ref{transitions},
illustrating the relative weakness of these lines in the HETG
spectrum. The most prominent candidate Fe~XVII line was
17.05~{\AA}. Its observed flux was comparable to the continuum level
there.) \citet{Rasmussen01} report a low iron-to-oxygen abundance
ratio and conclude that the Fe~L lines trace the swept-up ISM rather
than the ejecta.  This echoes the interpretation by \citet{Hayashi94}
that the iron emission is due to the forward shock interacting with
the ISM.

\subsection{Line Images} \label{Images}

\begin{figure}
\figurenum{3}
\plotone{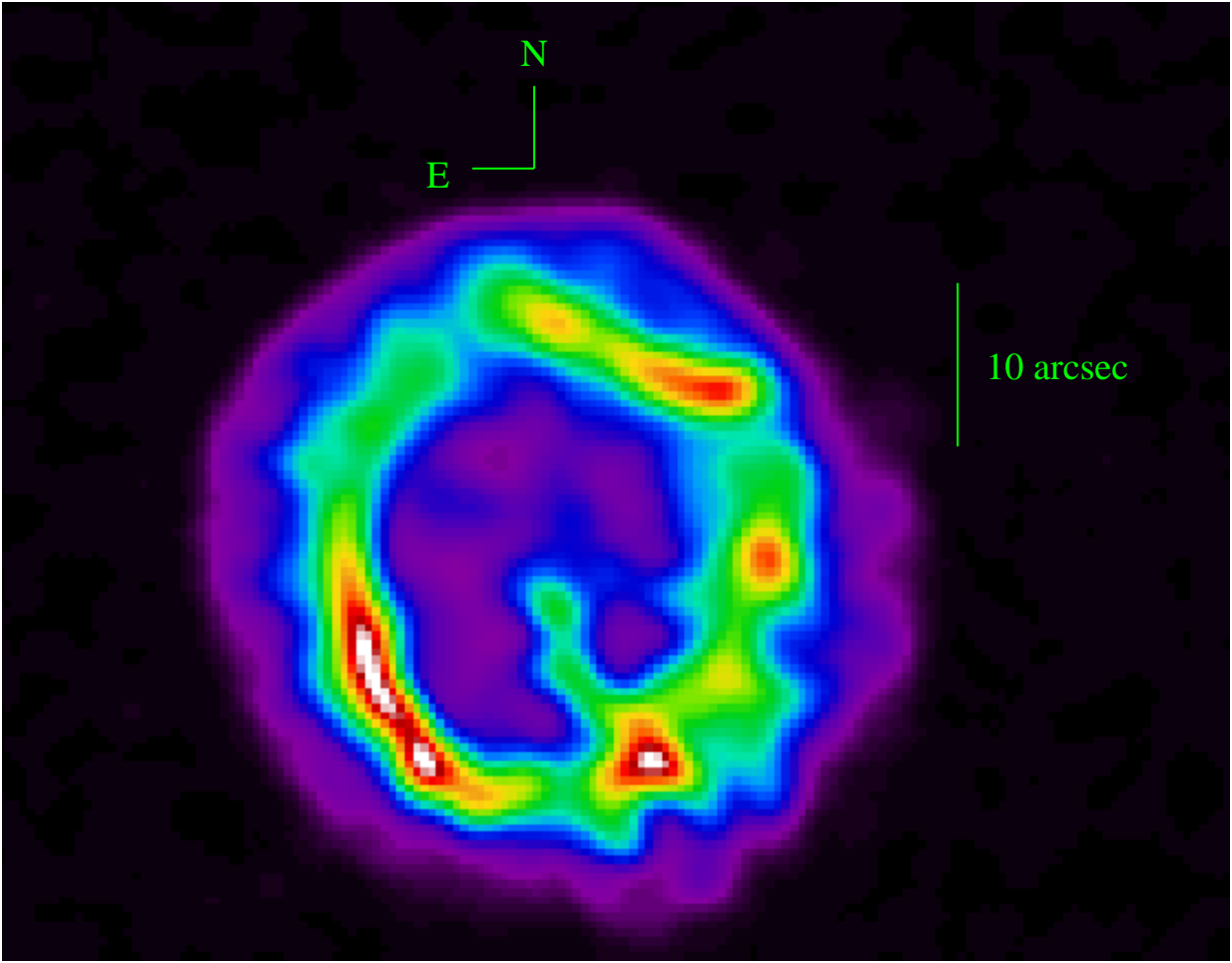} 
\figcaption[f3.eps]{Undispersed zeroth
order image of 1E\,0102-7219, with 2-pixel smoothing.
All energies have been included. In addition to the obvious shell
structure, note the bright ``shelf''-like feature at top, the bright
southeastern arc, and the radial ``spoke'' extending from a knot in
the southwest section of the ring.  The boundary of the blast wave is also
evident in the figure.} 
\label{zero}
\end{figure}

        The dispersed spectrum of E0102 shows the two-dimensional
structure of the SNR in several prominent lines. Figure~\ref{zero}
shows the undispersed zeroth order image of the SNR (all energies have
been included.)  The morphology of the remnant~\citep{Gaetz00} is that
of a ring with significant brightness variations.  The image shows a
bright knot in the southwest with a radial extension or ``spoke'', a
bright arc in the southeast, and a bright linear feature or ``shelf''
to the north. These features are visible to various degrees in the
dispersed X-ray line images. Also evident in the zeroth order image is the
boundary of the outward-moving blast wave noted by \citet{Gaetz00}.

\begin{figure}
\figurenum{4}
\epsscale{0.65}
\plotone{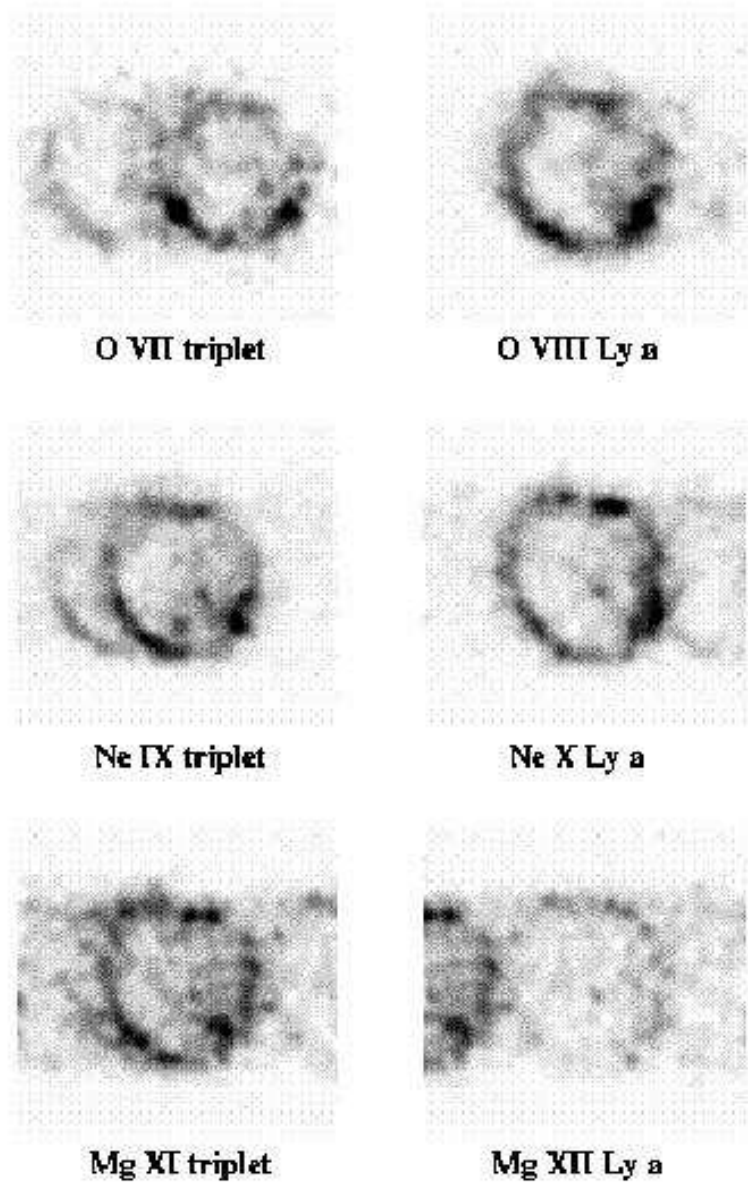} 
\figcaption[f4.eps]{The MEG~-1~order 
dispersed images formed by lines of oxygen (top), neon (middle) and
magnesium (bottom), shown in dispersion coordinates (rotated 
$\sim$17$^\circ$ with respect to the orientation in Figure~\ref{zero}).  For
each element, the helium-like n~=~2 to n~=~1 triplet is shown at left,
and the hydrogen-like n~=2~ to n~=~1 Lyman~$\alpha$ line is shown at right.
In each case, the forbidden line on the left of the triplet is bright
and easily distinguished from the resonance line on the right of the
triplet.  The intercombination line, located between, is very
faint. The faint ring to the right of Ne\,X~Lyman~$\alpha$ is
Ne\,IX~(1s--3p).  Note the disparity of ring size in the top row;
distortions of ring shape for Ne\,X~Lyman~$\alpha$ in the middle row; 
and absence of a complete ring in the southern portion of the hydrogen-like
stage of magnesium (bottom row).  These are taken to be indicators of the 
passage of the reverse shock, bulk matter Doppler shifts, and inhomogeneous 
plasma conditions, respectively. }
\label{ONeMg}
\epsscale{1.0}
\end{figure}

	Although the distinctive ring structure of the SNR is echoed
in each X-ray line, close inspection reveals differences line-by-line.
The dispersed images of prominent lines from the MEG~$-1$ order are
shown with a uniform spatial scale and 2-pixel Gaussian smoothing in
Figure~\ref{ONeMg}.
The helium-like triplet is shown to the left and the hydrogen-like
Lyman~$\alpha$ is to the right for oxygen (top), neon (middle) and
magnesium (bottom), respectively. 


	A comparison of the two oxygen images in the top row of
Figure~\ref{ONeMg} reveals a similar shape but a marked disparity in
ring diameter between the two ionization stages. This size difference
is taken to be an indicator of the progress of the reverse shock with
respect to the ejecta~\citep{Gaetz00, Flanagan01}, and is the subject
of Section~\ref{Shock}.  For both oxygen and neon, the forbidden
and resonance lines of the helium-like triplets are prominent and easily
resolved, providing potentially useful temperature diagnostics, as
discussed in Section~\ref{O7_ratios}.

	Distortions of the SNR ring shape can be seen by comparing the
ring from the Ne\,X~Lyman~$\alpha$ line, to the right in the middle row
of Figure~\ref{ONeMg}, with the undispersed image of Figure~\ref{zero}. 
In Section~\ref{Doppler} (Figure~\ref{triptych}), these images are compared
with the +1 order, providing compelling evidence for Doppler velocities of
the ejecta in the SNR.

	The two magnesium images in the bottom row of
Figure~\ref{ONeMg} show a striking difference in that the bottom
portion of the ring (i.e., the southeast arc that is so prominent in
other images) is virtually missing in the image formed by the
hydrogen-like Lyman~$\alpha$ line.  Given that the Mg\,XI triplet is
detected in the southern half (forming a complete ring), the lack of
the Mg\,XII line cannot be due to absence of the element magnesium,
but instead suggests insufficient ionization.  The signal-to-noise of
the Si\,XIII image was insufficient to include it in
Figure~\ref{ONeMg}.  Nevertheless, numerical comparison shows the
bottom portion suppressed relative to the top, analogous to the case
of Mg\,XII.  Thus, these images indicate the magnesium and silicon
plasma to be more highly ionized in the north relative to the south of
the SNR.  This is discussed more fully in Section~\ref{bad_plasma}.

\subsection{Flux Measurements} \label{Flux}

	The usual approach for measuring line fluxes involves folding
a spectral emission model through a response matrix and fitting
several components simultaneously.  However, there is so little
overlap in the dispersed line images of the {\it Chandra} HETGS
spectrum that we have adopted a more straightforward analysis of the
individual lines.  The measured values are given in
Table~\ref{tbl-flux} and the techniques are described below.


	The count rate for each bright X-ray line was obtained by using an
annular aperture that enclosed the observed ring. (The annular
aperture excluded much of the contribution by the ``spoke'',
representing typically less than 5\%. This was generally negligible
compared to the errors.  For measurements of the oxygen and neon
triplets, an elliptical aperture was
employed which captured the ``spoke'' contribution.) The background
was taken from regions immediately above and below the line of
interest. In both cases, an appropriate PI filter on ACIS energy was used
to minimize unwanted events.  For the line data presented here the
background was a small fraction of the total flux and was much smaller
than the error due to photon counting statistics.

	In some cases, the dispersed lines overlap other nearby lines
(i.e., the forbidden and resonance lines in Figure~\ref{ONeMg}), so
that assignment of events between the two X-ray lines is ambiguous.
In order to account for this, we extracted the events in a
sub-aperture devoid of overlap with the nearby contaminating line. The
measurement was then scaled up to obtain an estimate for the full
aperture. In making this correction, we took into account the fact
that the SNR brightness varies significantly around the ring. We
employed the zeroth order image filtered closely on the energy of
interest, and assumed that the distribution of events in the
sub-aperture relative to the full aperture was the same for the
dispersed ring as for this filtered zeroth order image.  
Accounting for surface brightness variations gives a result within
10\% of that derived assuming a uniform surface brightness.

	For each of the observation intervals, Obsid~120 and
Obsid~968, there are up to four independent flux measurements that may
be obtained: MEG +1 and $-1$ orders, and HEG +1 and $-1$ orders. Where
feasible, each of these measurements was made. The best counting
statistics were obtained with the MEG spectrum.  Nevertheless,
differences between raw measurements tended to be greater than
expected from counting statistics alone, and are attributed to
residual calibration effects.  In-flight calibration measurements
suggest systematic efficiency differences between backside-illuminated
and frontside-illuminated CCDs, and differences between HEG and MEG
grating measurements. To account for these calibration effects, we
have increased individual flux measurements of X-ray lines falling on
frontside CCDs by 4--19\%. (See
http://space.mit.edu/ASC/calib/hetgcal.html.   Backside CCD
measurements are believed to be correct, so that only frontside
measurements were adjusted in this way.)  We have also adjusted the
individual flux measurements of lines measured with HEG by 2--8\% to
normalize them to MEG measurements (see \citet{POG02},
Figure~8.26. Note that normalizing the HEG measurements to MEG is an
arbitrary choice, as it is currently unknown which represents the
correct standard.)  Since several such individual measurements are
used to determine the flux for each X-ray line reported in
Table~\ref{tbl-flux}, the impact of these corrections is effectively
smaller, typically ranging from 2--10\% for frontside CCD correction
and $\pm$2\% for HEG correction. The resultant overall correction
factors (weighted by source counts) are given in
Table~\ref{tbl-flux}. The largest calibration correction factor is
less than 10\%.

	To empirically estimate the continuum, we extracted regions of the
dispersed spectrum where X-ray line images were absent, and applied an
absorbed Bremsstrahlung fit. The best-fit model was then used to
estimate the continuum contribution for individual line energies
nearby. We found that line-free regions longward of the O\,VII triplet,
and between the O\,VII triplet and O\,VIII~Lyman~$\alpha$, could be 
fit and the continuum contribution in that region (i.e. for lines
below $\sim$~0.7~keV) could be estimated by an absorbed
bremsstrahlung model with k$T_e$ $\sim$0.9 and 
$N_H = 8\times10^{20}$~cm${}^{-2}$ (solar abundances).  
We found that the estimated continuum contribution
at the various oxygen line energies agreed well with continuum
estimates obtained from an independent fit to whole-spectrum ACIS data
by \citet{Plucinsky01}.  We took the error in our continuum estimates
to be 20\% based on typical differences among estimates obtained by
these and similar model fits.  (For the case of O\,VIII~Lyman~$\gamma$,
which lies rather far outside the energy range bracketed by the
line-free regions, we have enlarged the error to account for the
uncertainty in the model there.)  We estimate that the continuum
contributes less than 10\% to the measured flux at the bright lines of
the O\,VII triplet and O\,VIII~Lyman~$\alpha$, but more than 20\% of the
measurement for the weaker oxygen lines (i.e., O\,VII~(1s-3p),
O\,VIII~Lyman~$\beta$ and O\,VIII~Lyman~$\gamma$).  We could not obtain
a reliable continuum estimate at the energies of the neon, magnesium
and silicon lines, as the model we use in the oxygen line region
overpredicts the continuum at higher energies.  Further work is needed
to refine our estimates and to correct the other lines for their
continuum component.

	Table~\ref{tbl-flux} lists the mean values of the corrected
measurements for the set of lines bright enough to be measured by this
technique; the error bars in the table reflect the scatter among the
individual measurements in addition to the statistical errors.  Also
listed is the estimated contribution to this measurement due to the
continuum, and the net result after subtracting the continuum
component.

\subsection{Plasma Diagnostics} \label{Plasma}

High resolution grating spectra provide a means to probe plasmas using
individual X-ray lines. Ratios formed from different lines of the
same element serve as particularly useful plasma diagnostics because
they eliminate the impact of uncertainties in abundance or
distance. If lines from the same ion are selected, dependence on the
relative ionization fractions is reduced. (This provides, for
example, a useful diagnostic for electron temperature.)  By selecting
lines that are close in energy, the impact of uncertainties in the
column density is minimized.

\citet{Rasmussen01} reported the integrated high-resolution spectrum
of E0102 from the RGS of XMM--Newton, and employed several line ratios
as temperature diagnostics. In addition to the lines evident in the
HETGS spectrum, the RGS spectrum revealed lines of hydrogen-like
carbon, and marked the absence of nitrogen. (These lines fall outside
the HETG energy range.) 

Based on the images of Figure~\ref{ONeMg} and the discussion of
Section~\ref{Images}, the plasma conditions for magnesium and silicon
are distinctly different in the northern and southern portions of the
remnant.  Plasma diagnostics from ratios of integrated flux
measurements for these elements cannot be expected to give a
meaningful result.  However, the neon and oxygen images in
Figure~\ref{ONeMg} do not exhibit this obvious inhomogeneity, and
integrated fluxes might be applicable.  The disparity in oxygen ring
sizes seen in Figure~\ref{ONeMg} indicates that the emission regions
are different for the two ionization stages (as discussed in
Section~\ref{Shock Model}). However, a plasma model which accounts for
such an evolving ionization state can in principle accommodate this
radial dependence.  We have employed such a model to assess plasma
conditions based on oxygen line ratios.

Table~\ref{tbl-lineratios} lists useful diagnostic line ratios
obtained from oxygen fluxes in Table~\ref{tbl-flux}.  These have been
corrected for underlying continuum.  Ratios of neon and magnesium
lines are not included because of the uncertainties that remain in
estimating and subtracting the underlying continuum component.  The
raw ratio of observed fluxes (uncorrected for absorbing column) is
given in Table~\ref{tbl-lineratios}, followed by the ratio emitted at
the source assuming a column density of $N_H$ =
$8\times10^{20}$~cm${}^{-2}$ with solar abundances~\citep{Hayashi94,
Blair89}. This allows direct comparison with the XMM--Newton
results~\citep{Rasmussen01}.  The 90\% confidence contours are listed
in the last column.

We employed a non-equilibrium ionization collisional plasma model from
XSPEC version 11.1~\citep{Arnaud96} to calculate expected line
ratios. The model, vnpshock~\citep{Borkowski01}, is a plane-parallel
shock model that allows the user to select separate electron and ion
temperatures.  The ionization timescale, $\tau$, which defines the
progress of the plasma towards equilibrium, assumes a range of values
between lower and upper limits. (The evolution of the plasma is
mediated by collisions: $\tau$ equals the product of elapsed time and
electron density, and has units of s~cm${}^{-3}$.)  By incorporating a
range of $\tau$, the model takes into account the evolving nature of
the ionization. (We opted not to use another XSPEC non-equilibrium
ionization model, NEI, because it assumes a single fixed $\tau$.)  We
set $\tau_{\rm lower} = 0$, and additionally set the ion temperature
equal to the electron temperature. (We generally found the diagnostic
ratios were dictated by electron temperature, regardless of how the
energy was partitioned between the electrons and ions.)  We used
ISIS~\citep{Houck00} to run the XSPEC model and generate tables of
expected line ratios.  Figure~\ref{plasma} shows the ranges of $T_e$
and $\tau = \tau_{\rm upper}$ from the vnpshock model that are
compatible with the 90\% confidence limits for key line ratios listed
in Table~\ref{tbl-lineratios}.

\subsubsection{O\,VII line ratios} \label{O7_ratios}

	Ratios of lines that are close in energy are relatively
insensitive to the value of $N_H$. Thus, the lines of the O\,VII
triplet can provide particularly valuable diagnostics.  However, the
faint O\,VII~intercombination line flux cannot be measured by the
techniques of Section~\ref{Flux}, which complicates the analysis. We
fitted the components of the triplet using a technique described in
Section~\ref{Dewey}, to best determine the proportion of this line in
relation to the forbidden and resonance lines of the O\,VII triplet.
The results are shown in Table~\ref{tbl-dd} and corresponding ratios
are listed in Table~\ref{tbl-ddratio}.  The error for the fluxes
obtained by this fitting technique is estimated at $\sim$20\%. Within
these errors, the values agreed with the fluxes reported in
Table~\ref{tbl-flux} for the O\,VII resonance and forbidden lines.


	The HETGS O\,VII line ratios were found to provide a set of
self-consistent temperature diagnostics and are compatible with a
single temperature in the range of $\sim$0.14 -- 0.77~keV (i.e., 6.2
$<$ log $T_e$ $<$ 6.95).  The constraining ratio is shown in
Figure~\ref{plasma} as the forbidden-to-resonance ratio.
The remaining O\,VII ratios are compatible with this allowed region,
and do not further constrain it. Their contours are not plotted. The
electron temperature range allowed by the O\,VII ratios is in
agreement with the results obtained with XMM--Newton RGS spectra,
where O\,VII line ratios ($\alpha$$/$$\beta$ and $\beta$$/$$\gamma$
for helium-like oxygen;~\citet{Rasmussen01}) indicate a range of 0.35
$<$ $T_e$ $<$ 0.7 keV based on a
HULLAC~\citep{Bar-Shalom88,Klapisch77} model for an underionized
plasma.  

     The forbidden (f), intercombination (i) and resonance (r) lines
provide other useful diagnostics: R~=~f/i probes electron density, and
G~=~(i+f)/r can indicate departures from ionization
equilibrium~\citep{Pradhan82}.  The measured values for these ratios
are given in Table~\ref{tbl-ddratio}.  The corresponding XMM--Newton
measurements~\citep{Rasmussen01} are also listed for comparison.  Both
diagnostic ratios agree with XMM's observed values. R is consistent
with the value expected in the limit of low electron
density, and G is compatible with an ionizing plasma
(i.e., recombination suppressed) at a temperature of
$\sim$4$\times10^6$~K (log~$T_e$~=~6.6)~\citep{Pradhan82},
well within the allowed range indicated in Figure~\ref{plasma}.  

\subsubsection{O\,VIII line ratios} \label{O8_ratios}

The HETGS O\,VIII~line ratios are not very restrictive, establishing
only a lower limit on $T_e$.  For the ratio formed by the brightest
O\,VIII lines, Figure~\ref{plasma} (dashed line) indicates a
temperature above $\sim$0.14~keV (i.e. log $T_e$~=~6.2).  The best-fit
temperature for this ratio,
O\,VIII~Lyman~$\beta$/O\,VIII~Lyman~$\alpha$, is $\sim$0.36~keV.
Ratios formed with O\,VIII~Lyman~$\gamma$ were slightly more
constraining (i.e., T$_e$ $>$ 0.25~keV), but only the lower limit was
(marginally) allowable, and the best-fit value for the
O\,VIII~Lyman~$\gamma$/O\,VIII~Lyman~$\beta$ ratio was too high for
that contour to fall within the parameter ranges of
Figure~\ref{plasma}.  This contour is not plotted.  In the case of
XMM--Newton, the O\,VIII Lyman series ratios were found to be
anomalous.  (Lyman~$\beta$ flux was found to be higher relative to
Lyman~$\alpha$ than is predicted by electron impact ionization models.
This was also true for Lyman~$\gamma$ relative to Lyman~$\beta$, a
finding which is compatible with the HETGS result.)
\citet{Rasmussen01} considered the possibility of charge exchange as a
mechanism contributing to the observed ratios. Although all of the
HETGS O\,VIII Lyman series ratios are self-consistent and compatible
with a single ``allowed'' temperature range that overlaps that defined
by the O\,VII ratios, our errors provide little constraint and do not
contradict the XMM--Newton findings.

\subsubsection{O\,VII--O\,VIII line ratios} \label{O7_O8_ratios}

\begin{figure}
\figurenum{5}
\plotone{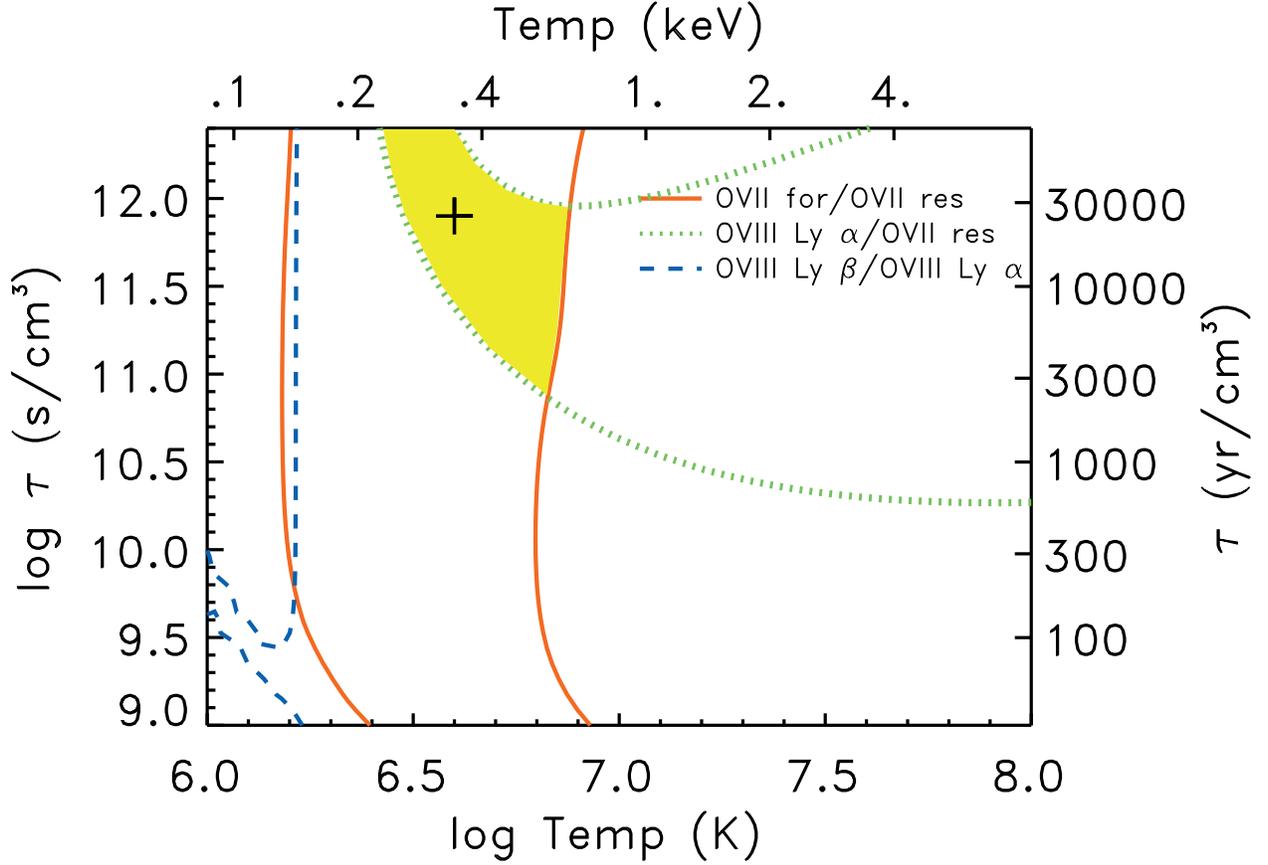}
\figcaption[f5.eps]{The 90\% confidence contours are plotted for
line ratios of the brightest measured O\,VII and O\,VIII lines, as
expected from a plane parallel shock model with input parameters of
$T_e$,~$\tau=\tau_{\rm upper}$.  The model (vnpshock) assumes
$T_e$~=~$T_i$, and a column density of $N_H = 8 \times 10^{20}cm^{-2}$
with solar abundances.  Some flux captured in the forbidden line
measurement is actually contributed by a nearby satellite line.  The
plotted contour accommodates this contribution.  The best-fit model is
marked by a cross.
\label{plasma}}
\end{figure}

	 Figure~\ref{plasma} illustrates that ratios between lines of
the same ion constrain the temperature but provide little information
with regard to the ionization timescale, $\tau$. Contours of the ratio
between the brightest lines of O\,VII and O\,VIII are indicated in the
figure and bring out the interplay between $\tau$ and $T_e$. There is
a subregion of $T_e$~--~$\tau$ parameter space (shaded yellow in
Figure~\ref{plasma}) that is compatible with all three of the contours
plotted: the region is delimited by 0.22--0.68~keV (6.4 $<$ $T_e$ $<$
6.9) and log $\tau$ $>$ 10.8~s~cm${}^{-3}$.  The best-fit values for
the $\beta$--to--$\alpha$ ratio for O\,VIII, the G-ratio for O\,VII,
and O\,VIII~Lyman~$\alpha$/O\,VII~Res are all consistent with an
oxygen plasma of temperature 0.34~keV (log~T$_e$~$\sim$6.6) and
log~$\tau$~$\sim$11.9~s~cm${}^{-3}$.  The HETGS best-fit model is
marked by a cross in Figure~\ref{plasma}, and the oxygen plasma model
is indicated in Table~\ref{model_params}.  The temperature
compares well with the best-fit value of 0.35~keV obtained from the
G-ratio measured by XMM~\citep{Rasmussen01}.

\subsubsection{Line ratios of other elements} \label{Ne_ratios}

We do not yet have an acceptable continuum estimate for the neon
energy region and therefore cannot obtain a ``clean'' line ratio
uncompromised by the continuum.  However, the continuum component had
only a slight (~$\sim$3\%) effect on the O\,VII f/r ratio (which
constrained the O\,VII plasma temperature).  If we neglect the
continuum contribution to analogous Ne\,IX triplet ratios, with the
vnpshock model we obtain a plasma temperature range of
$\sim$0.17--0.9~keV for Ne\,IX (assuming log $\tau$ $>$ 9.65
s~cm${}^{-3}$); this overlaps the temperature range found with the
XMM--Newton RGS.

\subsubsection{Caveat on Global Line Ratios}\label{bad_plasma}

	Some combinations of the bright lines can give a single plasma
fit~\citep{Davis01}, and we have seen that the set of oxygen lines
ratios is compatible with a single component plasma model.  However,
when the entire measurement set of Table~\ref{tbl-flux} is considered
(i.e., assuming no continuum correction for the neon measurements), a
single set of plasma parameters cannot adequately account for the set
of {\it Chandra} measurements of the integrated SNR spectrum for all
elements.  This echoes the conclusions reached by observers based on
CCD spectra.  \citet{Hayashi94} found from ASCA SIS data that a single
component plasma model was not acceptable for the global spectrum, and
fitted each element with its own non-equilibrium ionization plasma
model~\citep{HughesandHelfand85}.  They concluded that abundance
inhomogeneities exist in the plasma.  The insufficiency of a single
T$_e$~-~$\tau$ plasma to parameterize the global spectrum was
confirmed by \citet{Sasaki01} with XMM--Newton EPIC PN data, who
invoked two plasma components and additionally examined distinct
regions of the remnant.

	The {\it Chandra} HETGS spectrum clearly indicates that
spatially distinct plasma conditions exist within the SNR.  As shown
in Sections~\ref{Shock} and \ref{Shock Model}, there is spatial
separation of the helium-like lines from their hydrogen-like
counterparts.  This radial ionization structure of the SNR ring is not
the only indicator of plasma inhomogeneities.  As discussed in
Section~\ref{Images}, the lack of a complete ring in the southern
portion of the images for the Si\,XIII~resonance and
Mg\,XII~Lyman~$\alpha$ lines suggests higher ionization in the north
relative to the south.  Higher ionization can be achieved by a higher
temperature, or by a more advanced ionization timescale $\tau$. Since
the northern portion of the remnant displays a linear ``shelf''
appearance, it is conceivable that the expanding ejecta has
encountered a dense region of the circumstellar medium; the higher
density would encourage more rapid ionization.  \citet{Sasaki01}
examined EPIC~PN spectra from a northeastern segment (eastward of the
``shelf'') and from the bright X-ray arc in the southeast.  They found
that the northern spectrum had a higher fitted temperature and an
order of magnitude higher ionization timescale than the southern
region, indicating higher ionization.

	Given the evidence for spatially distinct plasma conditions,
it may be necessary to treat localized regions of the SNR
independently.  Although the O\,VII and O\,VIII global line ratios are
compatible with a single plasma model, it is possible (even likely)
that different plasma conditions dominate different parts of the
remnant. For example, the oxygen line ratios indicate a single
temperature, but this may be fortuitous -- it does not rule out the
possibility of multiple temperatures or local plasma variations.
Thus, any interpretation that relies on a global plasma model (as in
Section~\ref{Ejecta_mass}) must be treated with caution.  The
observations reported here do not have sufficient statistics to allow
plasma diagnostics for localized regions. However, a second GTO
observation of E0102 was carried out on December~20, 2002, providing
an additional 136~ks exposure time.  Future work will concentrate on
the ``shelf'' region and the bright southeastern arc, combining edge
profiles and local diagnostic line ratios to explore plasma
differences and model the plasma evolution in the wake of the reverse
shock.

\subsection{Elemental Mass Estimates} \label{Ejecta_mass}

	The best-fit model for the oxygen plasma may be used to
estimate the mass of oxygen in the X-ray emitting plasma.
The flux of a line observed at earth with no redshift or
column density, is given by

\begin{equation}
F~=~a_Z\frac{\epsilon(T_e)}{4\pi R^2} \int n_en_ZdV,
\end{equation}

\noindent where F is the flux in ph cm$^{-2}$s$^{-1}$, $a_Z$ is the 
abundance, $\epsilon(T_e)$ is the emissivity in ph cm$^3$ s$^{-1}$, 
R is the distance to the
source in cm and $n_e$ and $n_Z$ are the number densities in the
source of the electrons and the element Z, respectively.\\ 

For $\epsilon(T_e)$, we assume the emissivity generated by our
best-fit vnpshock plasma model (see Table~\ref{model_params}).  We
assume a distance, R, of 59~kpc~\citep{McNamara80}. We have measured
the flux, F, and assume a column density of
N$_H$=8$\times$10$^{20}$cm$^{-2}$ with solar abundances to obtain the
unabsorbed flux. To estimate the volume, we assume a simple geometric
ring-type model. \citet{Hughes88} found that an {\it Einstein} HRI
image of E0102-72 was well-fit by a thick ring. Our own analysis
(Section~\ref{Model}) suggests a similar geometry. Based on that
analysis, we can obtain a volume estimate: We assume a ring
represented by a portion ($\pm$30 degrees) of a shell of inner radius
3.9~pc (set by O\,VII edge profile measurements) and outer radius
5.5~pc. With that model, the volume estimate is
6.6$\times$10$^{57}$~cm$^3$. We assume a filling factor of 1.\\

        To determine the electron density, we have made the important
assumption that this is a pure metal plasma consisting of O, Ne, Mg,
Si and Fe. (Based on the dominance of the oxygen and neon lines and
the relative weakness of the iron lines and continuum in the spectrum,
we take the remnant to be ejecta-dominated and make the simplifying
assumption that the electrons are contributed predominantly by these
metals.) We assumed the oxygen, neon and magnesium were in
helium-like, hydrogen-like, or fully stripped configurations. We
further assumed the silicon is helium-like, and the iron is neon-like.
Finally, we assumed plasma conditions listed in
Table~\ref{model_params}, where the oxygen and neon plasmas are
characterized by the vnpshock models from HETG line diagnostics, and
the plasma models for magnesium, silicon and iron are based loosely on
\citet{Hayashi94}. \\

The assumption that the electrons are contributed by the metals gives
\begin{equation}
n_e = \Sigma_{Z}f_{Z}n_{Z}, 
\end{equation}
where $f_Z$ represents the number of electrons liberated per ion.
Multiplying both sides by $n_e$ and applying equation~1, we obtain

\begin{equation}
n_e^{2} = \Sigma_{Z}f_{Z}\left({4\pi R^2 F \over a_Z \epsilon(T_e) V}\right)
\end{equation}

	Using the measured fluxes of the brightest lines for these
elements, we obtained $n_e$ $\sim$ 0.9~cm$^{-3}$.  We find that oxygen
contributes about 69\% of the electron density, neon about 12\%, and
Fe, Mg, and Si contribute the remainder.  (The resultant value of
$n_e$ is not very sensitive to the specific model 
assumed for Fe, Mg and Si.) Since the contribution to $n_e$ by
hydrogen and other elements has been neglected, this estimate in
reality represents a lower limit to $n_e$.\\

	Substitution of n$_e$ into Equation~1 along with the measured
flux and emissivity from our best-fit oxygen plasma model yields an
estimate for the density of oxygen ions, n$_O$, from which we obtain
the mass of the oxygen ejecta.  The result, $\sim$6~M$_{\odot}$, is
listed in Table~\ref{model_params}. Similar analysis yields
$\sim$2~M$_{\odot}$ for the neon ejecta, but the plasma model is less
certain.\\

\begin{figure}
\figurenum{6}
\plotone{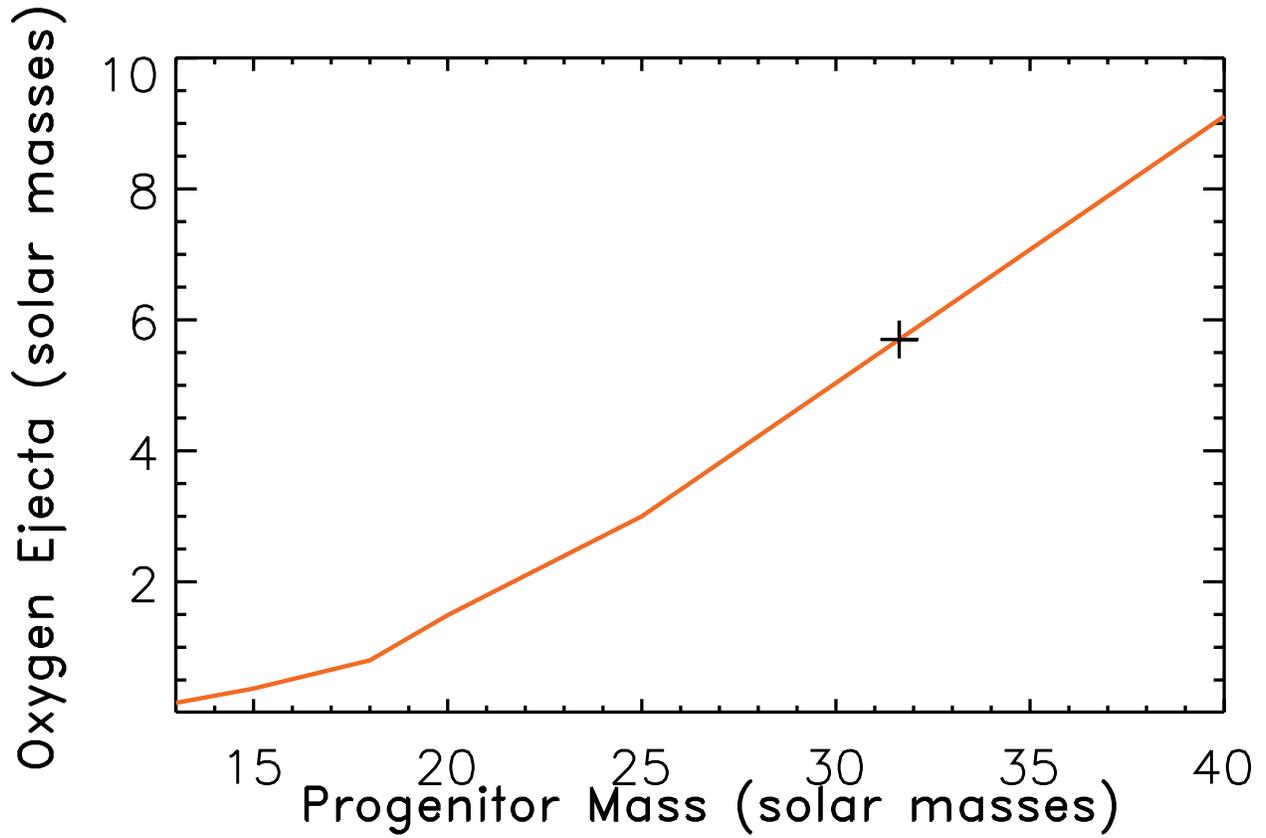} 
\figcaption[f6.eps]{Nomoto et al. 1997 predict specific
amounts of oxygen as a function of progenitor mass. The estimated
ejecta mass (marked by a cross on the plot)
indicates a massive progenitor of $\sim$32~M$_{\odot}$, assuming
a linear interpolation between the nearest models.
\label{nucleosyn}}
\end{figure}

        We used the nucleosynthesis models of \citet{Nomoto97} to
relate our estimate of oxygen ejecta mass to the progenitor mass.
Oxygen provides a particularly sensitive indicator of progenitor mass,
as shown in Figure~\ref{nucleosyn}. Our estimate of oxygen ejecta mass
suggests a massive progenitor of between Nomoto's 25~M$_{\odot}$ and
40~M$_{\odot}$ models.  The correlation between ejecta mass and
progenitor mass is essentially linear for Nomoto models from
25~M$_{\odot}$ to 70~M$_{\odot}$.  Assuming a linear interpolation
between models is appropriate, a progenitor mass of
$\sim$32~M$_{\odot}$ is indicated.\\


        \citet{Hughes94} has analyzed ROSAT HRI observations of
E0102-72.  He found evidence for a ring component and a shell
component, much as suggested by the HETG observation. Hughes finds
higher densities for the ring component, (n$\sim$6.0~cm$^{-3}$,
assuming solar abundances) and obtains a mass estimate of up to
75~M$_{\odot}$ for the x-ray emitting gas of the SNR (assuming a
filling factor of 1.)  He concludes that, even for a small filling
factor of $\sim$0.1, the progenitor was a massive star.\\

        \citet{Blair00} examined optical and UV spectra and compared
derived ejecta abundances to the models of \citet{Nomoto97}.  Their
abundance ratios appear to be well approximated by a 25~M$_{\odot}$
model.  Because they find no significant Fe or Si, they suggest that
the progenitor was a W/O star that exploded as a type Ib
supernova. Interestingly, they estimate approximately twice as much Ne
as predicted by the 25~M$_{\odot}$ model. The HETG estimate for the
neon ejecta, $\sim$2~M$_{\odot}$, is also larger than expected, about
three times what would be expected from 25~M$_{\odot}$ or 40~M$_{\odot}$
Nomoto models.\\

        The HETG results are consistent with \citet{Blair00} and
support the conclusion of a massive progenitor.  Several assumptions
have a significant impact on our calculations.  The most important is
the assumption of a ``pure metal'' plasma. If this assumption is
incorrect (and instead hydrogen dominates the plasma composition), the
value of n$_e$ will be underestimated by a factor of order $\sim$20.
The ejecta mass estimate is smaller by the same factor.  This would
place the progenitor mass in the range 13--15~M$_{\odot}$.  Our
assumptions about volume, filling factor $\eta$, and oxygen emissivity
$\epsilon$ are all less significant, with M$_O$~$\propto$~$\sqrt{\eta
V/ \epsilon}$. Even accommodating large uncertainties (i.e., a factor
of two in volume, $\eta$ as low as 0.1, and a factor of 2.5 variation
in $\epsilon$, accounting for the full allowed range in
Figure~\ref{plasma}), the results do not indicate a progenitor less
massive than 20~M$_{\odot}$. \\

	Assuming n$_e$$\sim$0.9~cm$^{-3}$, inspection of
Figure~\ref{plasma} implies that the ionization age of E0102 is longer
than the kinematic age (1000--2000 years).  Indeed, if the best-fit
$\tau$ ($\sim$25,000~yr~cm$^{-3}$) is correct, the kinematic age would
require n$_e$~=~12--25~cm$^{-3}$, although much smaller values are
needed (n$_e$~=~1--2~cm$^{-3}$) for the lower limit of $\tau$ in the
``allowed'' range of Figure~\ref{plasma}.  (For the XSPEC NEI model,
the emission-weighted $\tau$ is lower than for the vnpshock model: The
best-fit NEI value, $\tau$~$\sim$~4,500~y~cm$^{-3}$, would require
n$_e$~=~2--5~cm$^{-3}$ for an elapsed time of 1000--2000~yr.)  Given
that our estimate of n$_e$ obtained with a ``pure metal'' plasma model
represents a lower limit, these values are within the range of
uncertainties.

\section{Ring Diameter and Ionization} \label{Shock}

	The general ring structure of the SNR recurs in each of the images of
the individual X-ray lines in Figures~\ref{disperse} and
\ref{transitions}, but an important systematic difference is seen in
ring diameter.  This is evident in the top row of Figure~\ref{ONeMg},
which juxtaposes dispersed lines of helium-like and hydrogen-like
oxygen on the same spatial scale for comparison.  The helium-like line
is clearly emitted by a region of smaller diameter than the
hydrogen-like line.  A similar effect may be noted for neon (middle
row of Figure~\ref{ONeMg}).

\begin{figure}
\figurenum{7}
\plotone{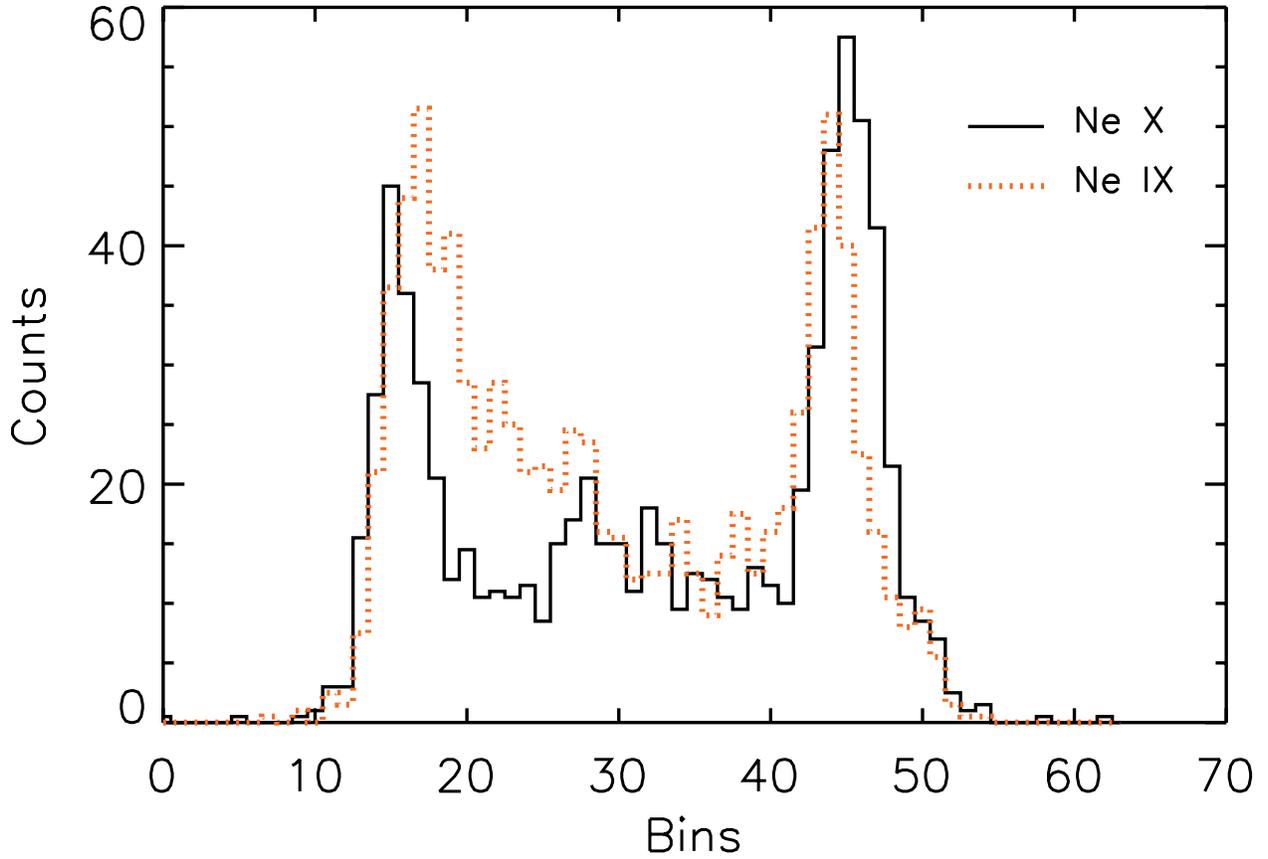}
\figcaption[f7.eps]{The cross dispersion histograms of the MEG $-1$ orders of
helium-like Ne\,IX and hydrogen-like Ne\,X lines are overlaid. The
difference in the two histograms shows that the emitting regions for
the two X-ray lines are different. The H-like Ne\,X  line is generated 
at larger radius, closer to the site of interaction between the CSM
and the ejecta. This suggests the action of the reverse shock. 
\label{crosscut}}
\end{figure}

The top and bottom edges of the bright ring images in the dispersed
spectrum were measured by extracting the intensity profiles along a
segment perpendicular to the dispersion direction. (The analysis was
restricted to perpendicular regions in order to minimize the effects
of Doppler shifts, which distort dispersed images along the
dispersion direction.)  The profiles, or cross-dispersion histograms,
for the neon lines of Figure~\ref{ONeMg} are shown in
Figure~\ref{crosscut}. The cross-dispersion direction is vertical in
each image of Figure~\ref{ONeMg}; thus, the left and right peaks of
the histogram in Figure~\ref{crosscut} respectively trace the
intensity profiles through the bright arc of the southeast and the
``shelf'' to the north.  Clearly, the helium-like neon ring is
measurably smaller than that of its hydrogen-like counterpart.

	
An empirical model was used to fit the emission profile and localize 
the peak of the emission. The model contained
the essential elements described in Section~\ref{Model}: an expanding,
ring-like shell inclined to the line-of-sight.  To better fit
the steep peak profile, the radial distribution was given a power law
dependence.

\begin{figure}
\figurenum{8}
\plotone{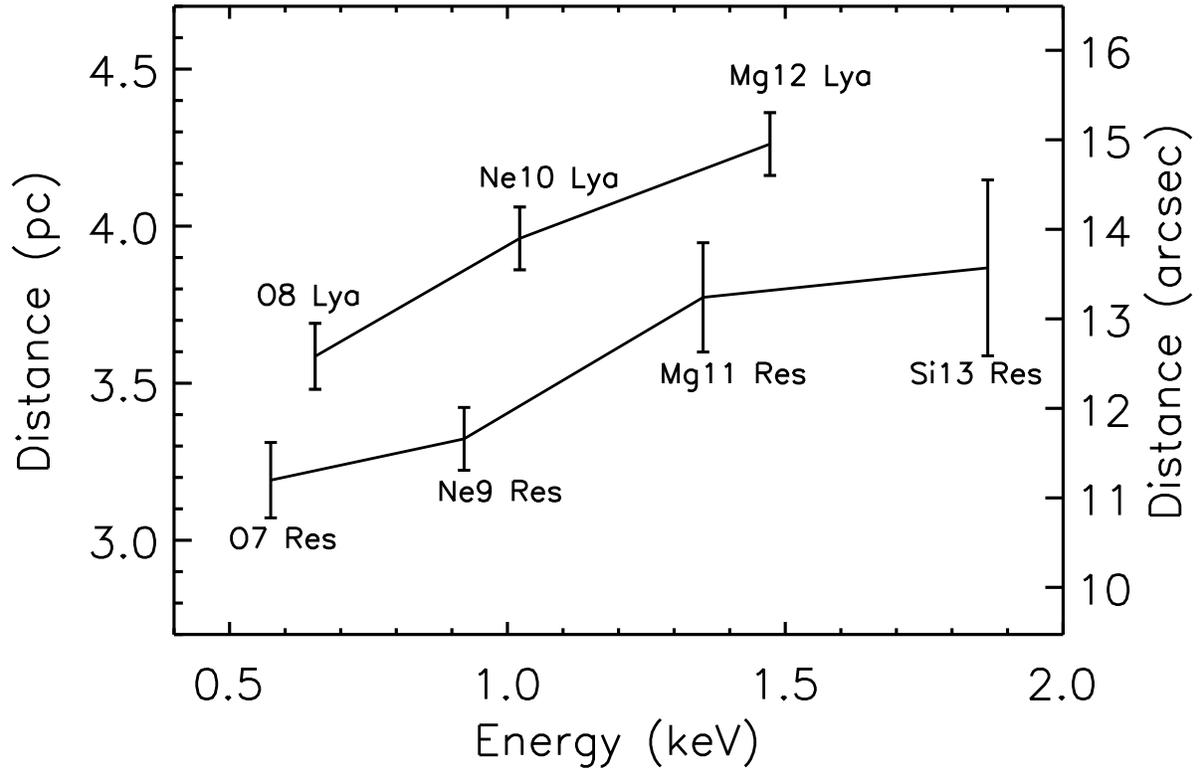}
\figcaption[f8.eps]{Measured radial distance of the northern
edge of the SNR (along the ``shelf'') corresponding to the brightest 
X-ray lines of 1E\,0102.2-7219.  The hydrogen-like
lines (upper curve) lie outside their helium-like counterparts 
(lower curve) for each element. Error bars on distance are dominated 
by the quality of the model fit which determines edge location. 
\label{measured_rings}}
\end{figure}

	The fitted location of the SNR edge along the northern
``shelf'' is plotted for seven bright X-ray lines (corresponding to
n=2 to n=1 transitions) in Figure~\ref{measured_rings}.  For each
element, the hydrogen-like lines (connected by the top curve) lie
outside their corresponding helium-like lines (connected by the bottom
curve) by a few arcseconds.  This is clearly an ionization effect,
unrelated to stratification or segregation of elements, because
different ionization stages of the {\it same} element are separated. 


	We attribute the spatial separation of ionization states to
changing ionization structure due to passage of the reverse shock
through the ejecta.  The reverse shock, driven by the retardation of
the ejecta as it sweeps up CSM, propagates inward relative to the
frame of the (moving) ejecta~\citep{McKee74}.  In this case, the
plasma in the outer regions of the ejecta has had a longer time to
react to the passage of the reverse shock and has experienced a higher
degree of ionization. This was the conclusion reached by
\citet{Gaetz00} based on an X-ray difference map and radial profile
analysis of O\,VII and O\,VIII emission from direct {\it Chandra} ACIS
images of E0102.  The HETGS spectral images clearly confirm this
ionization stage separation for each of the elements.

\subsection{Ionization Structure} \label{Shock Model}

\begin{figure}
\figurenum{9}
\plotone{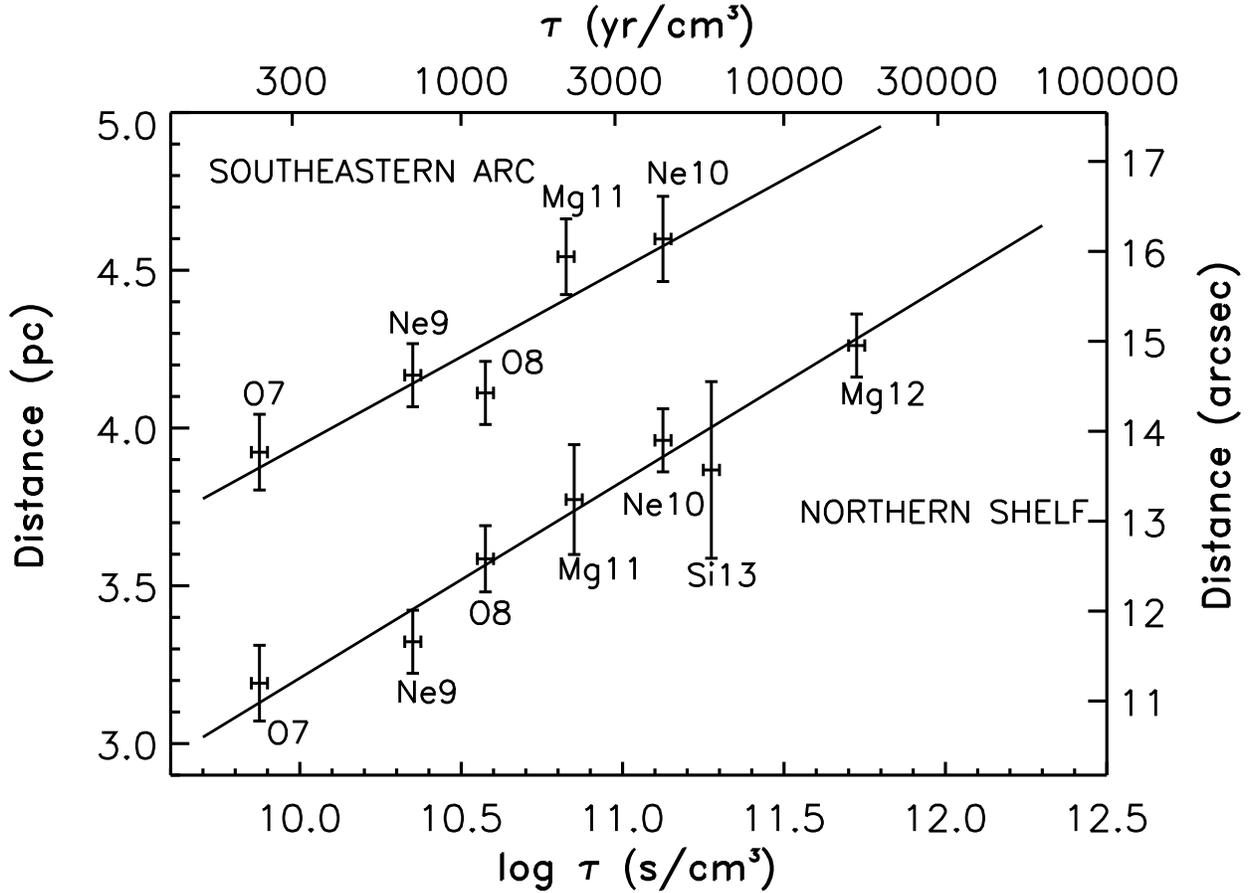}
\figcaption[f9.eps]{The ionization timescale ($\tau$) for which
each X-ray line exhibits maximum emissivity is plotted against the
measured radial distance, for two regions of the SNR: the linear
``shelf'' feature to the north, and the bright arc in the southeast.
A single constant electron temperature of 1.14~keV has been assumed
(Sasaki, {\it et al.} 2001). This plot is not expected to reflect the
actual specific conditions of the SNR plasma, but to illustrate the
correlation between ionization and radius.  The plot is consistent
with an interpretation of progressive ionization structure due to
passage of the reverse shock through the ejecta. Error bars on
distance are dominated by the quality of the model fit which
determines edge location. Error bars on $\tau$ are set equal to the
grid spacing chosen for the plane parallel shock model, vnpshock, and
$\tau$ represents $\tau=\tau_{\rm upper}$.
\label{dist}}
\end{figure}

        If the progress of the reverse shock is indeed the controlling
mechanism for the spatial segregation of Figure~\ref{measured_rings},
we hypothesize that the location of the SNR edge, (i.e., the peak of
the radial emission profile), correlates with the ionization
timescale, $\tau$.  We apply a simple model, assuming a fixed T$_e$ of
1.14~keV~\citep{Sasaki01} and the plane parallel vnpshock
model~\citep{Borkowski01} in a uniformly mixed plasma. For each X-ray
line, at a fixed temperature the emissivity reaches its peak at a
unique value of $\tau$.  We assign this value of $\tau$ to the X-ray
line, and plot it in Figure~\ref{dist} against the measured
radial distance of the edge (from the SNR center).  Measurements are
independently plotted for two edges of the SNR: the bright linear
``shelf'' to the north, and the southeastern arc.


	The general trend in Figure~\ref{dist} shows increasing
$\tau$ with increasing radial distance. The radial distribution
suggests that the various elements are intermingled, e.g., the bottom
curve (corresponding to the northern shelf) in
Figure~\ref{dist} shows lines of O\,VIII and Mg\,XI situated
between Ne\,IX and Ne\,X.  The similarity of the emission regions of
the three elements of Figure~\ref{ONeMg} is also compatible with an
assumption of substantial blending of the elements.  The ionization
stages of these elements are interleaved in just the order one would
expect from a homogeneous plasma with an inward-propagating shock.  The
specific values of $\tau$ depend on the assumed temperature in the
model, and indeed are selected for maximum emissivity.  These are not
expected to reflect the actual specific conditions of the SNR plasma,
but to illustrate the correlation between radius and ionization
timescale: The monotonic trend seen in Figure~\ref{dist} holds
for any temperature over a wide range (i.e., 0.5 to 1.5~keV). Although
any workable model must consider additional parameters, this
correlation is compatible with an interpretation in which the arcsec
differences in ring diameter are attributable to the changing
ionization structure resulting from the reverse shock.

\section{Image Distortions and Doppler Shifts} \label{Doppler}

	Doppler shifts due to center-of-mass bulk motion along the
line of sight will produce a systematic shift in the position of all of the
dispersed images proportional to wavelength. In contrast, high
velocity motions within the remnant (relative to its center of mass)
will cause Doppler shifts which {\it distort} the dispersed images along
the dispersion direction. (No distortions are introduced in the
cross-dispersion direction by velocities in the SNR). The magnitude of
the observed distortions therefore provides an emission-weighted
measure of the velocity structure in the X-ray emitting gas in the
remnant.

\begin{figure}
\figurenum{10}
\plotone{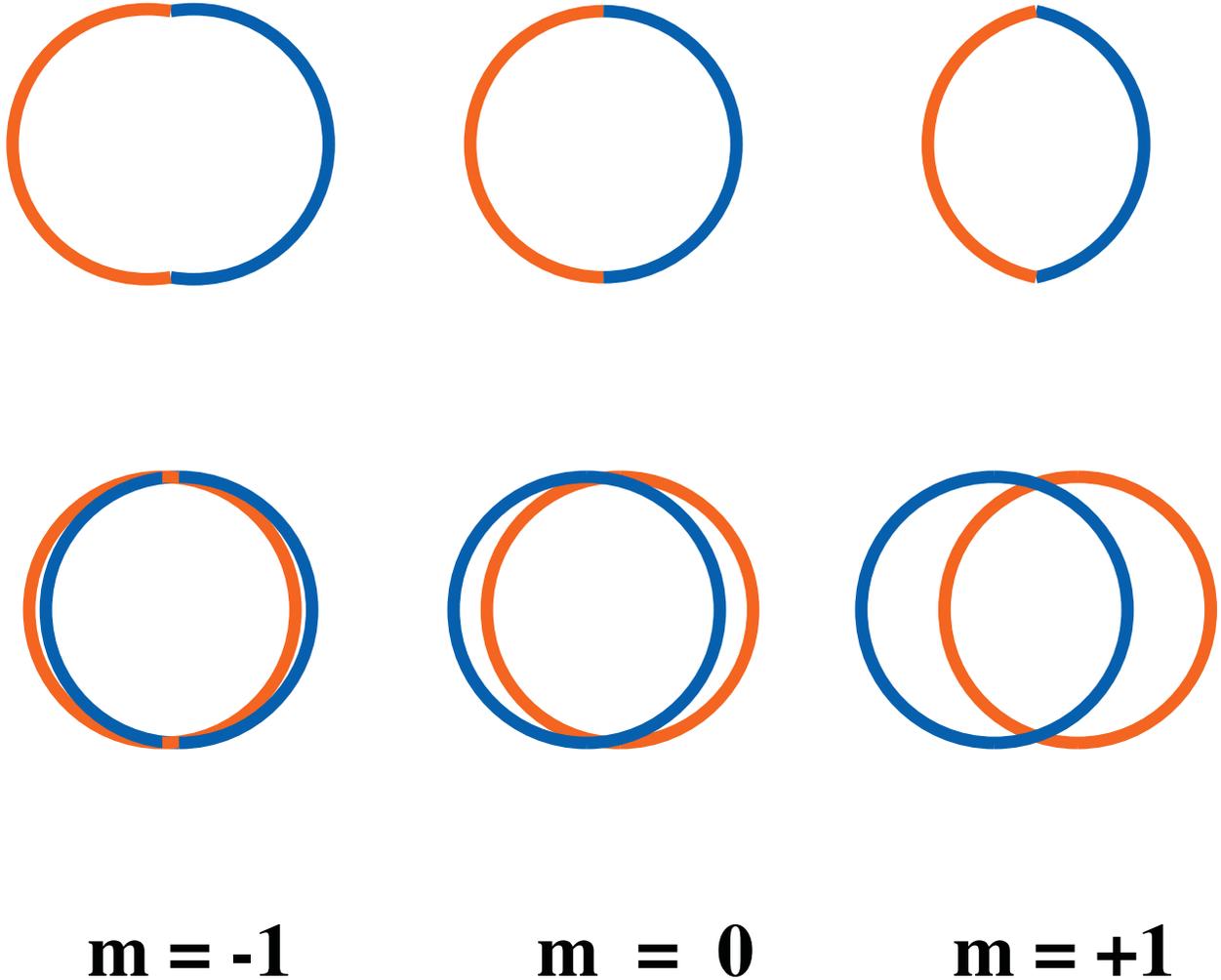}
\figcaption[f10.eps]{Cartoon illustrating that Doppler shifts
distort dispersed order images.  The top row (center) depicts a source
comprised of a thin circular ring with redshifted eastern half and
blueshifted western half. Doppler redshifts in the dispersed images
produce dispersion further outward, away from zeroth order, and
blueshifts are a little less dispersed, toward the zeroth
order at the center. This results in dispersed images which are not
circular, as shown by the ``stretched'' $-1$ order (left) and the
``squeezed'' +1 order (right).  The bottom row illustrates a situation
similar to our model for E0102.  The center panel schematically
illustrates a 3-dimensional source appearing as a thick ring with
separated red- and blueshifted components, such as found in
Figure~\ref{Doppler2D}.  (Such an arrangement might be associated with
an expanding non-spherical distribution of matter inclined to the
line-of-sight.)  For simplicity, only the inner and outer edges of
the ring are represented. Due to Doppler shifts, the $-1$ order
dispersed image (left) shows a narrowed ring, while the +1 order image
(ring) shows a broadened ring.  This effect is evident in
Figure~\ref{triptych}.  Thus, Doppler shifts result in +1 and $-1$
order images which differ in shape (top row), and effective width
(bottom row) of the ring.
\label{doppler_cartoon}}
\end{figure}

	Velocity structure can be distinguished from spatial
variations in the emissivity by combining information from the two
sets of gratings and from the plus and minus dispersion orders.  A
dispersed image in the light of a single line that is distorted due to
intrinsic spatial variations will look identical on either side of the
zeroth order (the plus and minus order dispersed images should look
the same).  A distortion due to a wavelength (Doppler) shift will
appear with opposite effects in the plus and minus orders, (i.e., a
shift to longer wavelength moves to the right in the plus order but to
the left in the minus order image, showing reflectional symmetry about
zeroth order), with the constraint that any distortions seen in the
MEG data should amplified by a factor of two in the HEG data due to
the difference in grating dispersion (with small corrections for the
different roll angles of the HEG and MEG dispersions).  This is
illustrated in Figure~\ref{doppler_cartoon}: A thin ring of emission
with red- and blueshifted regions will distort as shown in the top
row. A thick ring, as might be represented by an expanding cylinder
tilted to the line of sight, would show the effect in the bottom row.
The $-1$ order image appears sharper, while the +1 order is broadened.


		The impact of velocity structure is clearly evident in
the {\it Chandra} spectrum, as shown in Figure~\ref{triptych}.  This
figure displays side-by-side the dispersed MEG $-1$ order, the
undistorted zeroth order, and the MEG +1 order for
Ne\,X~Lyman~$\alpha$.  Overlaid on these images are alignment rings to
assist in identifying distortions.  There are clear distortions of the
dispersed images relative to the zeroth order. Moreover, the -1 and +1
order Ne\,X~Lyman~$\alpha$ images are distinctly different from each
other, with a sharper $-1$ order and a blurred or broadened +1 order,
analogous to the situation depicted in the bottom row of
Figure~\ref{doppler_cartoon}. These characteristics indicate velocity
structure within the SNR, and suggest a first approach to modeling
this structure. These topics are discussed in the remainder of this
section and in Section~\ref{Model}, respectively.


\subsection{Analysis} \label{Doppler anal}

	For the analysis of velocity structure, we focused on
Ne\,X~Ly~$\alpha$ and O\,VIII~Ly~$\alpha$ because these lines are
bright and their images have minimal contamination from nearby lines.
We relied on Obsid 120 alone for most of the analysis, adding Obsid
968 for confirmation.  Finally, we used event coordinates given in the
tangent-plane coordinate system (i.e., level~1 coordinates) to
facilitate direct comparison between dispersed and undispersed images
in the same coordinate system.

	To quantify the distortions in the dispersed images, we
extracted dispersed and undispersed images in such a way that
undistorted features would have identical pixel coordinates.  This is
illustrated in Figure~\ref{houck_anal}. First we selected a reference
pixel coordinate at the center of the zeroth order image.  We then
computed the coordinates of the reference point in the dispersed image
using the line wavelength, observation roll angle, and calibrated
values of the grating dispersion and tilt angle.  We extracted
first-order dispersed events centered on this computed position, and
formed an image with one axis parallel to the dispersion direction and
the other along the cross-dispersion direction.  By comparing the
images of the dispersed lines with undispersed zeroth-order images
filtered closely on the ACIS-S energy of the line, we were able to
measure the distortions at various positions along the ring, and
assign a corresponding Doppler velocity.

	Using this image alignment technique, the pixel coordinates of
the extracted images were aligned to within the uncertainty of the
grating wavelength scale, corresponding to a velocity uncertainty of
$\sim$200 km~s${}^{-1}$.  Because the extracted images were
accurately aligned using the rest wavelengths of the emission lines,
we conclude that the Doppler shift due to motion of the center of mass
of E0102 is below our detection limit, indicating that the
center-of-mass bulk velocity of the remnant is less than
$\sim$200~km~s${}^{-1}$.

\subsection{Doppler Shift Statistical Significance} \label{KS}

To establish Doppler shifts as the cause of the distortions, we
applied the Kolmogorov-Smirnov (KS) test to the event data to
investigate the possibility that the apparent distortions might be
consistent with Poisson noise.  We adapted it to compare event
distributions taken from slices in the various images both along the
dispersion direction and along the cross-dispersion direction.

{\it A priori}, if Doppler shifts are the cause of image distortions,
we expect different conclusions from the KS test depending on the the
orientation: Doppler shifts affect the dispersion direction, but do
not affect the cross-dispersion direction.  Using the language of the
KS test, we expect that the cross-dispersion slices are being drawn
from the same population, but, because of the Doppler distortions, the
dispersion-direction slices are not.  We conclude that, to high
confidence, the observed dispersion-direction distortions are real and
are not merely due to statistical fluctuations.

\subsection{Centroid Shifts} \label{Cen-Shift}

Having concluded that the distortions are real and attributable to
Doppler shifts, the next step is to quantify these distortions in
terms of the velocity structure in the SNR. In general, the line of
sight samples a distribution of velocities leading to Doppler shifts
which would smear the dispersed images in a way that is difficult to
quantify in terms of a simple statistic.  However, because of the high
degree of symmetry in E0102, these Doppler shifts combine somewhat
fortuitously to ``sharpen'' the SE limb of the $-1$ order dispersed
images and to shift the centroid of that limb in a way that is
consistent with bulk motion of the entire SE limb away from the
observer.

Although this localized centroid shift is consistent with bulk
motion, the overall structure can be more naturally explained in a
model based on radial expansion, similar to that depicted in
the bottom row of Figure~\ref{doppler_cartoon}. In this section,
we focus on those portions of the remnant where we observe
directly measurable Doppler shifts.  These cases are of interest
because they represent a direct, model-independent detection of
motion within the remnant. In Section~\ref{Dewey} we use a
Maximum-Entropy technique in which the dispersed and undispersed
images are fitted simultaneously to extract a detailed model of
the remnant's spatial structure and velocity field.  

Centroid shifts along the dispersion direction were determined by
measuring image centroids within an extraction box, adjusted so that
the result was insensitive to the details of the size and position of
the box.  Confidence limits were estimated through Monte-Carlo trials
using subsamples of observed events.  Our results for oxygen and neon
are summarized in Tables~\ref{tbl-o8shifts} and \ref{tbl-ne10shifts},
and discussed in Sections~\ref{O8_doppler} and~\ref{Ne10_doppler},
respectively.

Although measurement of the centroid shift has the advantage of
being model-independent, it is difficult to apply to regions with
complex morphology and reveals only the integrated properties of
the region, yielding an emission-weighted mean velocity. For these
reasons, the technique is not well suited to portions of the
remnant which are smeared in the dispersed image, i.e., the
western portion of the $-1$ order. For this reason,
Tables~\ref{tbl-o8shifts} and \ref{tbl-ne10shifts} report results
only for the eastern side of the SNR, and only a single mean
velocity component is obtained in each case.

\subsubsection{O\,VIII~Ly~$\alpha$} \label{O8_doppler}

The MEG $-1$ order image for O\,VIII~Ly~$\alpha$ allowed a clean
measurement for its eastern side, but the western side was blended
with a nearby X-ray line, O\,VII~(1s-3p).  Although the MEG +1 order
is unblended and could provide a cleaner measurement, it had
insufficient counts to be useful.  Thus, only eastern limb
measurements are reported in Table~\ref{tbl-o8shifts}.  The SE limb of
the MEG -1 dispersed O\,VIII~Ly~$\alpha$ image shows a large centroid
shift relative to the energy-filtered zero-order image.  It is
consistent with a recession velocity on the order of $\sim$1000 $\pm$
100 km s${}^{-1}$, indicating that the material dominating this
emission is probably on the back side of the remnant.

Although the zeroth order image is filtered around the energy
of the O\,VIII~Lyman~$\alpha$ line, it is contaminated with
O\,VII~triplet photons.  This could affect the reference zeroth order
position, with a resultant error in the Doppler shift measurement.
Moreover, corroboration of the Doppler shift interpretation using the
MEG +1~order image is not possible because the plus order image has
insufficient counts.  Due to blending and low surface brightness, we
were unable to determine whether or not the He-like triplet lines show
the same distortion.  The large velocities associated with the Doppler
shift measurements of the O\,VIII~Lyman~$\alpha$ line are comparable
with velocities measured for optical knots.  These knots generally lie
interior to the brightest portions of the X-ray bright ejecta, and
also show complex, asymmetric structure~\citep{Tuohy83}. 

\subsubsection{Ne\,X~Ly~$\alpha$} \label{Ne10_doppler}

The MEG $-1$ order image of Ne\,X~Ly~$\alpha$ had measurable red
shifts on the eastern side, but the dispersed images shown in
Figure~\ref{triptych} indicate that both red and blueshifts are
associated with the western side.  Because of the complex
morphology of the western side, no unambiguous centroid shifts
were apparent and therefore no measurements are reported in
Table~\ref{tbl-ne10shifts} for the western side of the SNR.  The
observed centroid shifts in the SE limb of the MEG $-1$ order
Ne\,X~Ly$\alpha$ image are significantly smaller than those found for
O\,VIII~Ly~$\alpha$, partly due to the shorter wavelength.  The
observed shifts are consistent with a recession velocity of $\sim$
450 $\pm$ 150 km s${}^{-1}$ in the SE limb. 

\subsubsection{Two Velocity Components} \label{red_blue}

The Doppler measurements of the SE limb revealed redshifts, suggesting
material on the backside of the remnant.  However, because the SE limb
of the MEG $-1$ order image is sharper than the +1 order image in both
O\,VIII and Ne\,X, we can conclude that {\it both} receding (red) and
approaching (blue) velocity components are present in that region. As
shown in Figure~\ref{doppler_cartoon}, this line of sight velocity
structure fortuitously ``sharpens'' the image in the -1 dispersed order
but ``blurs'' the +1 order.  Centroid shift measurements are inadequate
to describe this situation, where red and blue shifts coexist
within the same two-dimensional region of the
sky-image. Section~\ref{Dewey} describes the alternative approach we
have taken to map the velocity structure in two dimensions.

\subsection{Two Dimensional Spatial/Velocity Analysis} \label{Dewey}
 
To examine the SNR velocity structure in more detail, 
we have developed Doppler velocity imaging
techniques to extract a three-dimensional, spatial-velocity data cube
representation of the source, analogous to narrow-band Fabry-Perot
imaging in the optical/UV.  We began with a simple model of the SNR
with discrete velocity structure, as applied to the
Ne\,X~Lyman~$\alpha$ line.  This source model consisted of a simple
data cube with two spatial dimensions and a wavelength dimension.
Five wavelengths were selected distributed about the
Ne\,X~Lyman~$\alpha$ line rest wavelength (12.1322~\AA) with Doppler
shifts corresponding to velocities -2V,~-V,~0,~+V,~+2V. An additional
wavelength of 11.56~\AA~ was added to account for the rest wavelength
of the weaker Ne\,IX (1s-3p) line whose image intersects the
Ne\,X~Ly~$\alpha$ image (see Figure~\ref{ONeMg}).  The spatial
dimension representation was a $34\times34$ array of square cells,
each measuring three ACIS pixels on a side.  For each velocity plane,
this array of cells traced the Ne\,X~Lyman~$\alpha$ emission in the
undispersed zeroth order image.

        The $34\times34\times6$ data cube model was forward-folded to
create modeled images of the plus and minus order dispersed data.  In
an iterative conjugate gradient scheme, the model data cube values
were then adjusted to obtain the best fit between the modeled
dispersed data and the measured data.  The figure of merit for the
fitting was the sum of the $\chi^2$ measure of model and dispersed
data agreement and a negative entropy term measuring the deviation of
the zeroth-order image planes from the observed zeroth-order image
(i.e., there is a penalty term when the velocity plane deviates from
the zeroth order image.) This procedure was carried out with the
velocity parameter V set to values between 500~km~s${}^{-1}$ and
1250~km~s${}^{-1}$.  The overall best-fit was found for
V$\sim$900~km~s${}^{-1}$ and gave modeled MEG +1 and $-1$ order images
that were nearly identical to the observed images.

	The zeroth-order data cube values can be used to create a
color-coded velocity image of the Ne\,X~Lyman~$\alpha$ emission.  In
the image of Figure~\ref{Doppler2D}, red represents the sum of the two
redshifted planes (1800~km~s${}^{-1}$ and 900~km~s${}^{-1}$), which
were added because each traced essentially the same spatial region;
green represents the $-$900~km~s${}^{-1}$ blueshifted plane, and blue
corresponds to the -1800~km~s${}^{-1}$ highly blueshifted plane.  The
zero-velocity plane is not shown, but lies roughly between the red- and
blueshifted regions.  This image clearly shows the spatial separation
and structure of the red- and blueshifted velocity components in the
remnant.  This figure is a preliminary result from this new analysis
technique and further refinement and error estimation is ongoing.


It is clear from the Figure~\ref{Doppler2D} that both the eastern and
western sides of the SNR contain red- and blueshifted components.
Along both sides of the SNR, the redshifted regions are situated
westward of blueshifted regions, echoing the arrangement depicted in
the bottom row of Figure~\ref{doppler_cartoon}.  The image suggests
spatially offset rings of red- and blueshifted emission, as might
occur for a cylindrical or elongated distribution which is viewed
off-axis.  The placement of the zero velocity emission is consistent
with such a model. In particular, the zero-velocity component does not
lie outside the red/blue regions, as would be expected for a spherical
distribution.  Such a simple model invites extending the
two-dimensional Doppler map to a three-dimensional physical picture of
the SNR, as discussed in the next section.

	The ``spoke'' region in Figure~\ref{Doppler2D} shows a
dominance of blue shifts, suggesting that this feature is on the front
(near side) of the SNR.  Fabry-Perot measurements by~\citet{Eriksen01}
also reveal blueshifts among [OIII] filaments measured in that region,
with maximum velocities in excess of 2100~km/s.  They conclude that
this may be a dense clump of shocked material on the front of the
remnant, although a component near zero velocity is also detected in
the spoke region. The \citet{Eriksen01} [OIII] measurements also
indicate more moderate blueshifts coincident with the SE arc seen in
the X-ray. \citet{Tuohy83} present [OIII] images of E0102 which
confirm the blueshifts seen in regions of the spoke and the SE arc,
and also report a redshifted filament nestled between.

	Our velocity plane analysis gives an estimate of the rms
velocity dispersion of the Ne\,X line of $\sim$1100~km~s${}^{-1}$.
From the broadening of emission lines observed with the RGS on {\it
XMM-Newton}, \citet{Rasmussen01} place a lower limit of
1350~km~s${}^{-1}$ on the expansion velocity of the ejecta assuming a
flat velocity profile for an expanding shell.  The rms equivalent for
this lower limit and profile is 800~km~s${}^{-1}$, in good agreement
with our velocity plane analysis.

\section{3D Spatial/Velocity Model} \label{Model}

\begin{figure}
\figurenum{11} 
\plotone{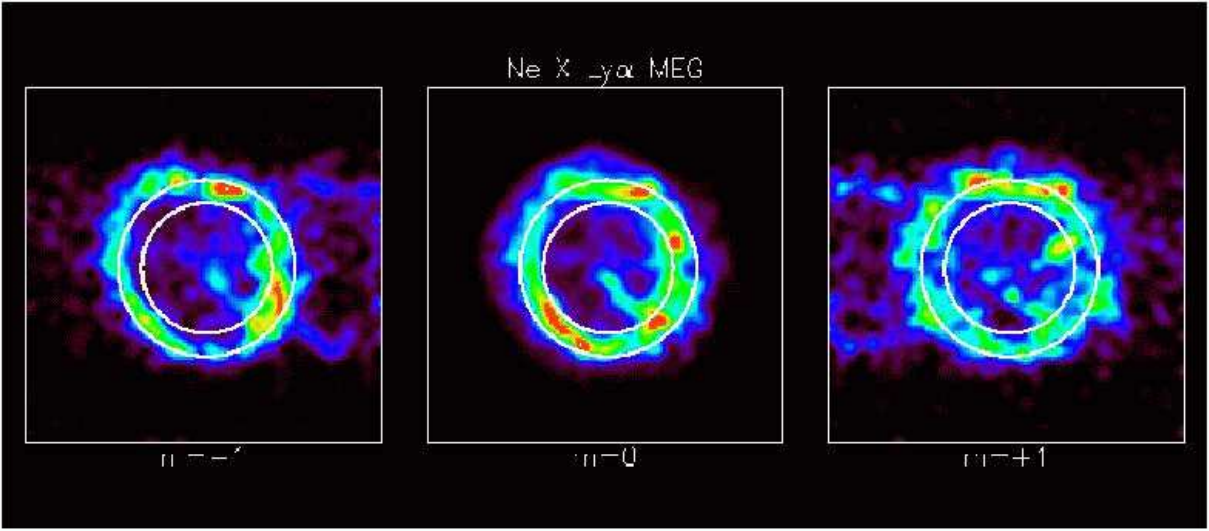} 
\figcaption[f11.eps]{Distortion is evident when dispersed images of
Ne~X~Lyman~$\alpha$ are compared with the zeroth order (filtered on
the energy of the line).  The leftmost panel shows the MEG -1~order
ring, the middle panel is the zeroth order, and the right panel is the
MEG +1~order ring. Overlaid on these images are alignment rings to
assist in identifying distortions. The sharpness of the -1~order ring
relative to +1~order indicates the presence of both red- and
blueshifts.
\label{triptych}}
\end{figure}

\begin{figure}
\figurenum{12} 
\plotone{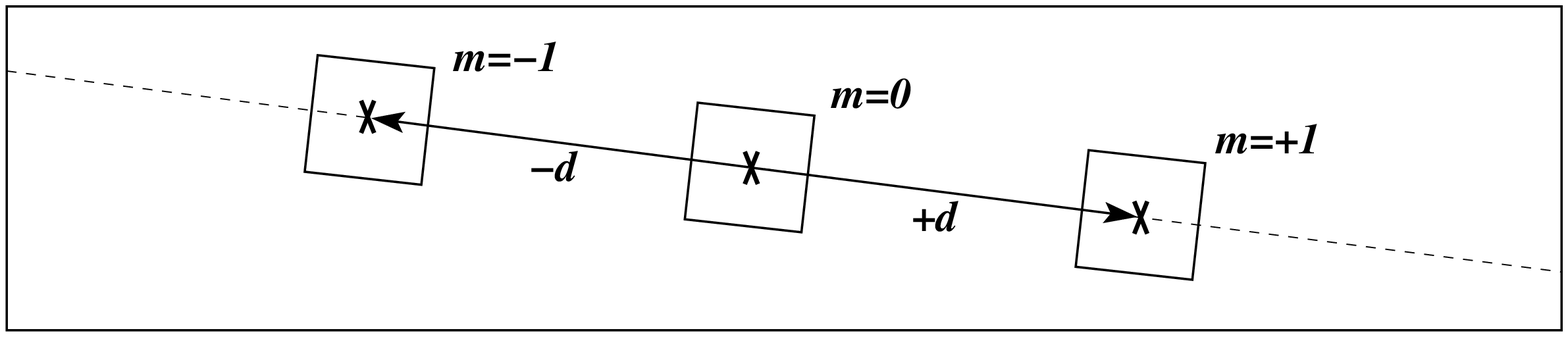} 
\figcaption[f12.eps]{Illustration showing how dispersed and
undispersed images are obtained in such a way that undistorted
features have identical pixel coordinates.  A reference pixel (X) is
selected at the center of the zeroth order image and a rectangular
extraction region of the desired size is centered on this reference
pixel.  For the wavelength of interest, the dispersed coordinates of
the reference pixel are computed, establishing the location of
corresponding extraction region for the dispersed image. First-order
events are extracted and images are formed with one axis parallel to
the dispersion direction and the other axis along the cross-dispersion
direction.
\label{houck_anal}}
\end{figure}

\begin{figure}

\figurenum{13}
\plotone{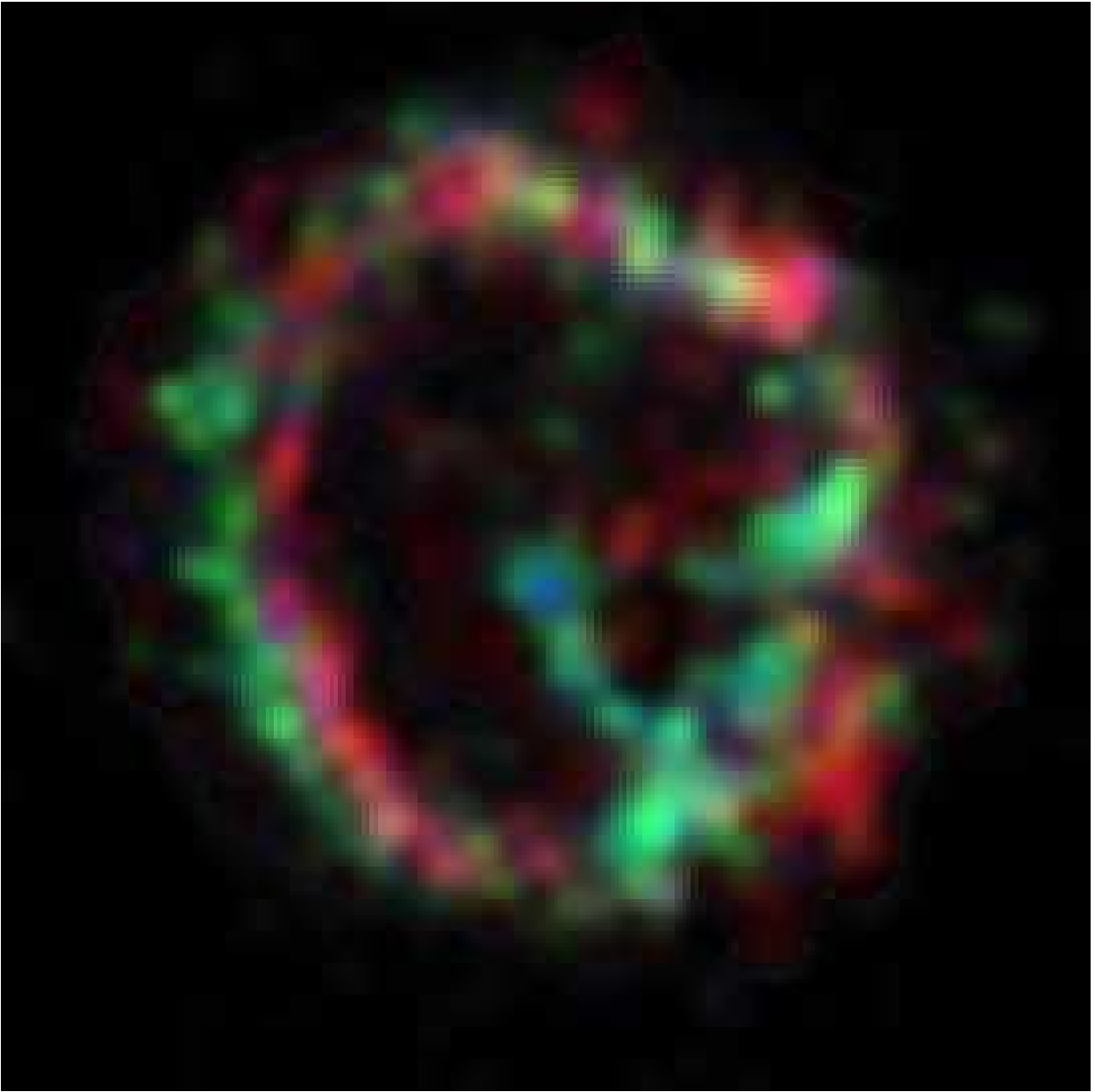}
\figcaption[f13.eps]{Velocity planes
referenced to the zeroth order image are depicted for
Ne~X~Lyman~$\alpha$.  A single red plane represents the sum of the two
redshifted planes +1800~km~s$^{-1}$ and +900~km~s${^-1}$, green is
the -900~km~s$^{-1}$ blueshifted plane, and blue corresponds to the
-1800~km~s$^{-1}$ blueshifted plane. The zero-velocity plane is not
shown, but lies roughly between the red- and blueshifted regions.
These three planes, normalized to their maxima, determine the RGB
contributions to the color map using the CXC/Ciao tool
``dm2img''. Thus, where redshifts and blueshifts coincide, the color
representation is yellow.  The striking red/blue offset is
qualitatively consistent with preliminary modeling, where the
emission is concentrated toward an equatorial plane inclined to the line of
sight. 
\label{Doppler2D}}
\end{figure}

\begin{figure}

\figurenum{14} 
\hspace*{2in}\psfig{file=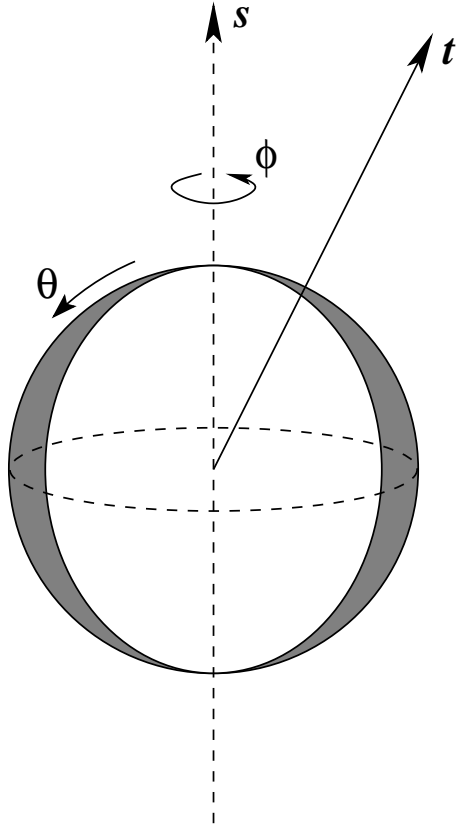,width=6cm}
\figcaption[f14.eps]{Schematic diagram of axisymmetric model
geometry. $\bf s$ is a unit vector along the axis of symmetry.  $\bf
t$ is a unit vector pointing toward the telescope.  The polar
coordinate is $\theta$ and azimuthal coordinate is $\phi$.  In the
preferred model orientation, $\bf t \cdot s = \cos 34\arcdeg$.  Viewed
from this angle, the matter distribution in projection approximates
the observed ellipticity of the zero-order image.  The major axis of
this ellipse is parallel to the vector $\bf t \times s$, which points
$\sim$10$^{\circ}$ clockwise from North. The emitting material fills a
spherical shell with inner radius $\sim$4.3~pc and thickness
$\sim$1.2~pc (these are merely representative numbers derived from the
zeroth order image). The density distribution within the shell is
indicated qualitatively by the thickness of the grey shaded region
(see text for details).  The emission fills the shell, but is
concentrated toward the equatorial plane.  This simple model
reproduces many of the features of the HETGS spectrum, but does not
reproduce the separation of redshifted and blueshifted components seen
in Figure~\ref{Doppler2D}.
\label{model_pic}
}
\end{figure}

We have constructed a preliminary empirical model of the
spatial and velocity structure.  Diffuse emission from the
interior of the observed ring rules out a purely toroidal
geometry, but is much weaker than would be expected from a
uniform spherical shell.  We infer an intermediate geometry
consisting of a non-uniform spherical shell with azimuthal
symmetry whose axis is inclined to the line of sight. This is
illustrated in Figure~\ref{model_pic}.  We take the inclination
to be such that the east side of the SNR is redshifted (i.e.,
tilted away from the observer) and the west side is
blueshifted.  The remnant may be intrinsically elliptical, but
a viewing asymmetry must nonetheless be present, as indicated
by the blurred +1 order and sharp -1 order of
Figure~\ref{triptych} and the suggestion of
Figure~\ref{Doppler2D}. 
Our empirical model of the emissivity 
distribution has the form
\begin{equation}
\varepsilon(r, \theta) = 
\varepsilon_0 f(r) \exp\left(-\frac{\cos^2\theta}{2\sigma^2}\right)
\end{equation}
where $r$ is the radial coordinate, $\theta$ is the polar
angle and where the radial distribution within the shell
has the form
\begin{equation}
 f(r) = \begin{cases}
    \left(\frac{r-r_{\rm min}}{r_{\rm max} - r_{\rm min}}\right)^\alpha &
 (r_{\rm min} \leq r < r_{\rm max}),\\
     0 & {\rm otherwise}
     \end{cases}
\end{equation}
The Gaussian $\sigma$ is chosen so that 85\% of the emission
originates within $\pm$30$^{\circ}$ of the equator. Within the
shell, we assume that the ejecta expands with velocity
proportional to radius.  Interestingly, \citet{Hughes88} deduced
that the surface brightness distribution measured with the {\it
Einstein} HRI also indicated that the emission was concentrated
in a thick ring rather than a spherical shell, although the
width of his ring was larger, distributed through an opening
half-angle of 67$^{\circ}$.  \citet{Hughes94} found that ROSAT
HRI data required both ring-like and shell-like components.  

Our simple 3D model for E0102 reproduces many of the features
of the HETGS spectrum. The expansion of the shell and inclination of
the symmetry axis to the line of sight effectively reproduce the
intriguing difference between the narrow $-1$ order and the broadened
+1 order seen in the dispersed images of Ne\,X~Lyman~$\alpha$ in
Figure~\ref{triptych}.  Moreover, inclining the model ejecta ring to
the line of sight results in enhanced emission due to projection
effects (i.e. limb brightening), and may partially account for the
enhanced emission in the vicinity of the northern ``shelf'' and the
bright southeastern arc.  Finally, the elements of this model provide
a good representation for the edge profiles.  However, this simple
model falls short in several respects. It does not correctly reproduce
the elliptical shape of the zeroth order image, and neglects
individual features such as the radial ``spoke''.  More importantly,
this model does not reproduce the striking separation of blueshifted
and redshifted components seen in Figure~\ref{Doppler2D} and depicted
for the m=0 order in the bottom row of Figure~\ref{doppler_cartoon}.

Other 3D models are being considered to describe the spatial and
velocity structure of the SNR.  A cylindrical or elongated (e.g. barrel-shaped)
distribution is an inviting candidate as it will make a better match
with the Doppler map of Figure~\ref{Doppler2D}.  In such a case, the
ring thickness seen in Figure~\ref{zero} would be interpreted as the
projection of a much thinner region extending in three-dimensions and
inclined to the line of sight. Such a model must simultaneously
satisfy the edge profiles and other constraints.  The second GTO
observation of E0102 taken in December, 2002, will help resolve these
questions. It was carried out at a different roll angle, and therefore
provides complementary Doppler information to the first
observations. The two sets of GTO observations, taken together, will
be used to refine the model.  Although the simple candidate 3D models
we have described are incomplete, they will serve as a point of
departure for future work.

\section{Discussion} \label{Discussion}

Our Doppler analysis of E0102 indicates velocities of order
$\sim$1000~km~s${}^{-1}$ and a toroidal (or possibly cylindrical)
distribution.  Several other oxygen-rich SNRs show evidence for a
ring-like geometry with expansion velocities
$\sim$2000~km~s${}^{-1}$. X-ray observations of Cas A by
\citet{Markert83} using the Focal Plane Crystal Spectrometer on the
{\it Einstein Observatory} revealed evidence for asymmetric Doppler
shifts which they interpreted as a ring of material with an expansion
velocity in excess of 2000~km~s${}^{-1}$.  (~\citet{Hwang01} consider
a highly asymmetric explosion and subsequent evolution to account for
their {\it Chandra} Doppler map of Cas~A.)  X-ray observations of
G292.0+1.8 by \citet{Tuohy82} revealed a bar-like feature which they
attributed to a ring of oxygen-rich material ejected into the
equatorial plane.  Examination of a velocity map generated from
optical observations of E0102-72 led \citet{Tuohy83} to suggest that
its velocity structure could be modeled in terms of a severely
distorted ring of ejecta. \citet{Lasker80} also found toroidal
expansion in the optical lines of N132D.

A cylindrical or toroidal distribution of ejecta can arise as a result
of the core collapse process or through the interaction of the
CSM. \citet{Khokhlov99} discusses evidence that the core collapse
process is asymmetric. Spectra of core collapse supernovae are
polarized, indicating an asymmetric distribution of ejected
matter~\citep{Wang01}. Neutron stars are formed with high space
velocities~\citep{Strom95} and ``bullets'' of ejecta penetrate
remnant boundaries~\citep{Fesen96,Taylor93}. Simulations of
jet-induced core collapse supernova explosions~\citep{Khokhlov99}
result in high velocity polar jets and a slower, oblate distribution of
ejecta. Two dimensional models of standing accretion shocks in core
collapse supernovae~\citep{Blondin03} are unstable to small
perturbations to a spherical shock front and result in a bipolar
accretion shock, followed by an expanding aspherical blast wave.

Asymmetry in the supernova remnant can also be imparted through the
influence of the CSM.  \citet{Blondin96} find that SNR show evidence
of being ``relics of supernovae interacting with non spherically
symmetric surroundings.''  \citet{Igumenshchev92} find aspherical
cylindrical symmetry in $\sim$30\% of remnants observed with
Einstein~\citep{Seward90}. One of the explanations they propose is
that the SNe explode in a medium with a disk-like density distribution
determined by the stellar wind from the precursor red supergiant. This
density distribution is believed to be heavily concentrated in the
equatorial plane. This produces a remnant elongated along the polar
axis and having lower density in that direction.  Such CSM density
distributions from red giant and red supergiants are expected to be
common, as evidenced by the case of planetary nebulae
(PN). \citet{Zuckerman86} found in a sample of 108 planetary nebulae,
about 50\% were bipolar. PN shapes are currently modeled by a fast
wind from a central star colliding with a cylindrically symmetric
stellar wind from the precursor red giant \citep{Kwok78,Kahn85}; where
mass-loss is enhanced towards the equator, elongation along the poles
results. \citet{Blondin96} have studied interactions of supernovae
with an axisymmetric CSM and find that the asymmetry generally follows
the asymmetry of the CSM~\citep{Blondin01}, but for a sufficient
angular density gradient, it is possible to obtain protrusions or
jet-like structures along the axis.  Thus, the asymmetric structures
inferred from our data, whether emission dominated by an equatorial
ring, or an elongated cylindrical-type structure, are compatible with
current models of SNe explosions and subsequent interactions with the
CSM.

The ionization structure of the SNR as revealed in Figure~\ref{dist}
provides a compelling picture of the action of the reverse shock.  For
two different interaction sites in the SNR (the northern ``shelf'' and
the SE arc), up to seven different X-ray lines follow a pattern that
can be simply explained by a time-dependent ionization in a uniform
abundance plasma. There appears to be substantial mixing of the
elements, as indicated by the remarkable similarity of the O, Ne and
Mg rings shown in Figure~\ref{ONeMg}, and the intermingling of
peak emission regions for lines of oxygen, neon and magnesium
as shown in Figure~\ref{dist}.

\section{Summary} \label{Summary}

	The HETGS spectrum of E0102 yields monochromatic images of the
SNR in X-ray lines of hydrogen-like and helium-like oxygen, neon,
magnesium and silicon with little iron. Plasma diagnostics using
O\,VII and O\,VIII lines summmed over the entire remnant confirm that
that the SNR is an ionizing plasma in the low density limit, and give
a best-fit oxygen plasma model represented by a plane-parallel shock
(vnpshock) of electron temperature $T_e$~=~0.34~keV and ionization
timescale log~$\tau$~$\sim$11.9~s~cm${}^{-3}$.  Assuming a pure metal
plasma, the derived oxygen ejecta mass is $\sim$6~M$_{\odot}$,
consistent with a massive progenitor.

	The dispersed X-ray images reveal a systematic variation of
ring diameter with ionization state for all elements.  We find that
this structure is consistent with the evolution of the ejecta's plasma
after passage of the reverse shock, and cannot be explained by radial
stratification of the elements. Future work will include a second
observation and focus on the ionization structure and emission
distribution in the northern linear ``shelf'' feature, and across the
bright southeastern arc.

	Distortions within the dispersed images reveal Doppler shifts
due to bulk motion within the remnant. Measurements for lines of neon
and oxygen indicate velocities on the order of 1000~km~s${}^{-1}$,
consistent with velocities measured for optical filaments. A
two-dimensional spatial/velocity map has been constructed which shows
a striking spatial separation of redshifted and blueshifted regions.

	A simple three-dimensional physical model of the SNR shows
rough agreement with many of the features attributed to Doppler
shifts.  This 3D model consists of a nonuniform spherical shell with
azimuthal symmetry, and with emission concentrated toward the
equatorial plane.  The symmetry axis is inclined with respect to the
line-of-sight, and the shell is expanding.  We find that this model
reproduces some, but not all, of the essential features of the HETGS
spectrum. Cylindrical distributions for the reverse-shocked ejecta
plus a blastwave component will be another candidate model for future
investigation.  The asymmetric structures implied by the data are
compatible with current models of SNe explosions and subsequent
interactions with the CSM.

\section*{Acknowledgments}

	We thank our anonymous referee for gracious and helpful comments.
We thank Glenn Allen, Norbert Schulz, Tom Pannuti and Herman
L. Marshall for helpful discussions. We thank Kazimierz Borkowski for
help in interpreting O\,VII data. We thank Kris Eriksen for helpful
discussions and providing a detailed look at his [OIII] Fabry-Perot
measurements. We thank Stephen S. Murray for reviewing and commenting
on the text.  We are grateful to the CXC group at MIT for their
assistance in analysis of the data.  This work was prepared under NASA
contracts NAS8-38249 and NAS8-01129, and SAO SV1-61010.

\clearpage

\input{tab1}

\input{tab2}

\input{tab3}

\input{tab4}

\input{tab5}

\input{tab6}

\input{tab7}

\input{tab8}

\clearpage



\clearpage

\end{document}

%% file: tab1.tex
\begin{deluxetable}{lcc}
\tabletypesize{\scriptsize}
\tablecaption{HETG Observations of 1E0102.2-7219 \label{tbl-obs}}
\tablewidth{0pt}
\tablehead{
\colhead{ } & \colhead{Obsid 120}  & \colhead{Obsid 968} 
}
\startdata
nominal observation time & 90 ksec & 48.6 ksec  \\
roll angle (degrees)\tablenotemark{a}	 & 12.0088 & 11.1762   \\
effective observation time & 86.9 ksec & 48.6 ksec  \\
date observed	& Sept 28, 1999	& October 8, 1999   \\
zeroth order \tablenotemark{b}	& 55.3 & 29.5 \\
MEG $\pm$ 1 order\tablenotemark{b}	& 45.8 & 25.5 \\
HEG $\pm$ 1 order\tablenotemark{b}	& 15.0 & 7.9 \\
 \enddata
\tablenotetext{a}{rotation angle about the viewing axis; positive is west of north}
\tablenotetext{b}{$10^3$ raw counts}

\end{deluxetable}

%% file: tab2.tex

\begin{deluxetable}{llrrrrrrr}
\tabletypesize{\scriptsize}
\tablecaption{Observed Flux \label{tbl-flux}}
\tablewidth{0pt}
\tablehead{
\colhead{X-Ray line} & \colhead{Energy}  & $\lambda$& 
\colhead{flux\tablenotemark{a}} &
\colhead{error\tablenotemark{a}} & \colhead{correction} &
\colhead{continuum\tablenotemark{a,c}} & \colhead{net flux\tablenotemark{a}} 
& \colhead{net error\tablenotemark{a}}\\
\colhead{} & \colhead{(keV)}  & \colhead($\AA)${} & \colhead{} & 
\colhead{(1$\sigma$)} &\colhead{factor\tablenotemark{b}}&\colhead{} & \colhead{} 
}
\startdata
O~VII~Forbidden & 0.5610 & 22.10 &13.95\tablenotemark{d,l} & 2.17 & 1.024 & .92 & 13.03 & 2.18\\
O~VII~Resonance & 0.5740 & 21.60 &24.38\tablenotemark{d} & 3.80 & 1.027 & .96 & 23.42 & 3.80\\
O~VII~Triplet & 0.5675 & 21.85 & 49.63\tablenotemark{d,l} & 1.54 & 1.027 & 2.54 & 47.09 & 1.62 \\
O~VIII~Lyman~$\alpha$ & 0.6536 & 18.97 & 37.31\tablenotemark{e} & 8.45 & 1.020 & 1.22 & 36.09 & 8.45\\
O~VII~1s--3p & 0.6655  & 18.63 & 6.23\tablenotemark{d}  & 1.31 & 1.040 & 1.29 & 4.94 & 1.34 \\
O~VII~1s--4p & 0.6977 &  17.77 & 2.97\tablenotemark{d}  &  .41 & 1.049 & 1.22 & 1.75 & 0.48 \\
Fe~XVII~NeVIII & 0.7270 & 17.05 & 2.89\tablenotemark{d}  &  .40 & 1.048  & 1.44 & 1.45  &  0.49\\
O~VIII~Lyman~$\beta$ & 0.7744 & 16.01 & 6.37\tablenotemark{f} & .67 & 1.079 & 1.28 & 5.09 & 0.72 \\
O~VIII~Lyman~$\gamma$ & 0.8168 & 15.18 & 3.75\tablenotemark{d} & .39 & 1.092 & 1.32\tablenotemark{k} & 2.43 & 0.66\tablenotemark{k}\\
Ne~IX~Forbidden &  0.9050 & 13.70 & 7.14\tablenotemark{i,l} & 1.24 & 1.097 &\\
Ne~IX~(Res~+~Intercomb) & 0.9225  & 13.49 & 13.49\tablenotemark{h} & 1.41 & 1.073 &\\
Ne~IX~Triplet & 0.9137 & 13.57 & 23.18\tablenotemark{i,l} & 1.28 & 1.092 &\\
Ne~X~Lyman~$\alpha$ & 1.0221 & 12.13 & 9.72\tablenotemark{h} & .66 & 1.048 &\\
Ne~IX~1s--3p & 1.0725 & 11.56 & 3.09\tablenotemark{g} & .24 & 1.046 &\\
Ne~IX~1s--4p & 1.1271 &  11.00 & 1.99\tablenotemark{g} & .39 & 1.045 &\\
Ne~X~Lyman~$\beta$ & 1.2108 & 10.24 & 2.07\tablenotemark{g} & .31 & 1.037 &\\
Mg~XI~Triplet & 1.3521 & 9.17 & 3.77\tablenotemark{j,l} & .27 & 1.024 &\\
Mg~XII~Lyman~$\alpha$ & 1.4725 & 8.42 & 1.01\tablenotemark{h} & .14 & 1.028 &\\
Si~XIII~Triplet & 1.8644 & 6.65 & 1.06\tablenotemark{h,l} & .20 & 1.037 &\\
 \enddata

\tablenotetext{a}{$10^{-4}$ ph~cm${}^{-2}$~s${}^{-1}$}
\tablenotetext{b}{Raw flux was multiplied by this factor to obtain
final flux; normalizes HEG to MEG measurements, and frontside to
backside CCD measurements} 
\tablenotetext{c}{20\% error assumed unless otherwise indicated} 
\tablenotetext{d}{from 4 MEG measurements}
\tablenotetext{e}{from 3 MEG measurements and 1 HEG measurement}
\tablenotetext{f}{from 3 MEG measurements} 
\tablenotetext{g}{from 4 MEG measurements and 2 HEG measurements} 
\tablenotetext{h}{from 4 MEG measurements and 4 HEG measurements} 
\tablenotetext{i}{from 4 MEG and 3 HEG measurements} 
\tablenotetext{j}{from 3 MEG and 4 HEG measurements} 
\tablenotetext{k}{40\% error assumed for continuum}
\tablenotetext{l}{includes satellite line contribution}

\end{deluxetable}

%% file: tab3.tex

\begin{deluxetable}{llcclll}
\tabletypesize{\scriptsize}
\tablecaption{Oxygen Line Ratios \label{tbl-lineratios}}
\tablewidth{0pt}
\tablehead{
\colhead{Lines} & \colhead{Ratio\tablenotemark{a}} & \colhead{error} & 
\colhead{Ratio\tablenotemark{b}} & \colhead{error} & \colhead{90\% contours\tablenotemark{b}}\\
\colhead{} & \colhead{} & \colhead{(1$\sigma$)} & \colhead{} & \colhead{(1$\sigma$)} & \colhead{}
}

\startdata
O~VIII~Lyman~$\alpha$ / O~VII~Resonance & 1.54 & .44 & 1.26 & .36 &  .67 -- 1.85 \\
O~VII~Forbidden / OVII~Resonance & .56 & .13 & .58 & .14 &  .35 -- .81\tablenotemark{c} \\
O~VIII~Lyman~$\beta$ / O~VIII~Lyman~$\alpha$ & .14 & .04 & .12  & .03 & .07 -- .18 \\
O~VIII~Lyman~$\gamma$ / O~VIII~Lyman~$\beta$ & .48 & .15 & .46  & .14 & .23 -- .69\\
O~VII~1s--3p / O~VII Resonance & .21 & .07 & .17 & .05 & .08 -- .26 \\
O~VII~1s--4p / O~VII~1s--3p & .35 & .14 & .34 & .13 & .13 -- .55 \\

 \enddata

\tablenotetext{a}{ Raw ratio: column density $N_H = 0$ }
\tablenotetext{b}{ Assumes column density $N_H = 8\times10^{20}$~cm${}^{-2}$ and solar abundances}
\tablenotetext{c}{ Contours 0.26--0.81 have been plotted in Figure~\ref{plasma} to accomodate the results of Table~\ref{tbl-ddratio}}

\end{deluxetable}


%% file: tab4.tex
\begin{deluxetable}{llrcrc}
\tabletypesize{\scriptsize}
\tablecaption{O~VII Triplet Fitted Line Fluxes \label{tbl-dd}}
\tablewidth{0pt}
\tablehead{
\colhead{Line} & \colhead{Wavelength} & \colhead{Flux\tablenotemark{a}} & \colhead{error} & \colhead{Flux\tablenotemark{b}} & \colhead{error}
}
\startdata
O~VII~Resonance & 21.6015 & 24.513 & 4.903 & 47.871 & 9.574 \\
O~VII~Intercombination & 20.8036 & 2.844 & 0.570 & 5.647 & 1.129 \\
O~VII~Forbidden & 22.0977 & 11.145 & 2.229 & 22.741 & 4.548 \\
\enddata

\tablenotetext{a}{ $10^{-4}$ ph cm$^{-2}$ s$^{-1}$, no $N_{H}$ correction}
\tablenotetext{b}{ $10^{-4}$ ph cm$^{-2}$ s$^{-1}$, $N_H = 8\times10^{20}$~cm${}^{-2}$ }

\end{deluxetable}

%% file: tab5.tex
\begin{deluxetable}{llrcrcc}
\tabletypesize{\scriptsize}
\tablecaption{O~VII Triplet Fitted Diagnostic Line Ratios \label{tbl-ddratio}}
\tablewidth{0pt}
\tablehead{
\colhead{Diagnostic} & \colhead{Ratio\tablenotemark{a}} 
& \colhead{error} 
& \colhead{Ratio\tablenotemark{b}} & \colhead{error} 
& \colhead{XMM--Newton\tablenotemark{c}} & \colhead{error}\\
\colhead{} & \colhead{} & \colhead{(1$\sigma$)} & \colhead{} & 
\colhead{(1$\sigma$)} & \colhead{} & \colhead{} 
}
\startdata
R~=~f/i & 3.9 & 1.1 & 4.0 & 1.1 & 3.4 & 0.6 \\
G~=~(i+f)/r & 0.57 & 0.15 & 0.59 & 0.15 & 0.55 & 0.03 \\
f/r & 0.45 & 0.13 & 0.48 & 0.13 &  &  \\
\enddata

\tablenotetext{a}{ raw ratio: $N_{H}$~=~0}
\tablenotetext{b}{ Assumes $N_H = 8\times10^{20}$~cm${}^{-2}$ and solar abundances}
\tablenotetext{c}{ \citet{Rasmussen01}, which assumes $N_H = 8\times10^{20}$~cm${}^{-2}$ and solar abundances}

\end{deluxetable}

%% file: tab6.tex
\begin{deluxetable}{ccccc}
\tabletypesize{\scriptsize}
\tablecaption{Assumed Plasma Parameters and Estimated Ejecta Masses \label{model_params}}
\tablewidth{0pt}

\tablehead{
\colhead{Element} &  \colhead{$T_e$} & \colhead{log $\tau$} &
\colhead{Plasma Model Basis} & \colhead{Ejecta Mass} \\
\colhead{} & \colhead{(keV)} & \colhead{(s~cm${}^{-3}$)} & \colhead{} 
& \colhead{(M$_{\odot}$)}
}
\startdata

Oxygen & 0.34\tablenotemark{a} & 11.9\tablenotemark{b} & HETG plasma diagnostics\tablenotemark{c} & 5.7  \\ 
Neon & 0.58\tablenotemark{d} & 11.9\tablenotemark{e}& HETG plasma 
diagnostics\tablenotemark{f}& 2.2 \\ 
Magnesium & 0.5 & 12.0 &  Hayashi {\it et al.} (1994) \\ 
Silicon & 0.6 & 11.80 & Hayashi {\it et al.} (1994) \\ 
Iron  & 3.25 & 10.45 &  Hayashi {\it et al.} (1994)\\ 

\enddata 

\tablenotetext{a}{Allowed range 0.22--0.68~keV.}

\tablenotetext{b}{Allowed range log~$\tau$ $>$ 10.8~s~cm${}^{-3}$.}

\tablenotetext{c}{See allowed region in Figure~\ref{plasma}.  Best-fit
values are discussed in Section~\ref{O7_O8_ratios}.}

\tablenotetext{d}{Allowed range 0.45--0.75~keV.}

\tablenotetext{e}{Allowed range log~$\tau$ $>$ 11.5~s~cm${}^{-3}$.
$\tau$=$\tau_{upper}$ for vnpshock model, as in Figure~\ref{plasma}
and Section\ref{Plasma}.}

\tablenotetext{f}{Plasma parameters were selected to represent the
``allowed'' region expected with vnpshock model and two line ratios:
Ne\,IX Forbidden/(Res+Intercomb), Ne\,X Lyman~$\alpha$/ Ne\,IX (Res +
Intercomb). The continuum component was not subtracted in considering
these ratios, and this model is less certain than the oxygen plasma
model.}

\end{deluxetable}

%% file: tab7.tex
\begin{deluxetable}{crr}
\tabletypesize{\scriptsize}
\tablewidth{0pt}
\tablecaption{Eastern Limb, O\,VIII Ly$\alpha$\label{tbl-o8shifts}}
\tablehead{
\colhead{Cross-Disp. Slice\tablenotemark{a}} & 
\colhead{Centroid Shift\tablenotemark{b}} & 
\colhead{Velocity\tablenotemark{c}}
}

 \startdata
 ( 20, 30)  &  $10 \pm  4 $ &  $1800 \pm 700$ \\
 ( 10, 20)  & $3.2 \pm 0.7$ &  $ 560 \pm 120$ \\
 (  0, 10)  & $3.0 \pm 0.5$ &  $ 530 \pm  80$ \\
 (-10,  0)  & $3.8 \pm 0.6$ &  $ 660 \pm 100$ \\
 (-20,-10)  & $6.2 \pm 0.4$ &  $1100 \pm  70$ \\
 (-30,-20)  & $3.1 \pm 0.8$ &  $ 550 \pm 140$ \\
  \enddata
\tablenotetext{a} {Range of cross-dispersion coordinates within the 
extracted image, in units of ACIS pixels. ACIS pixel size is 0.49 arcsec.}
\tablenotetext{b} {Shift of MEG -1 order relative to zeroth order, in units
of ACIS pixels. Reference point (0,0) in the zero-order image was 
J2000 (RA, DEC)= (16.00885,-72.03126).}
\tablenotetext{c} {km s${}^{-1}$}
\end{deluxetable}

%% file: tab8.tex
\begin{deluxetable}{crr}
\tabletypesize{\scriptsize}
\tablewidth{0pt}
\tablecaption{Eastern Limb, Ne\,X Ly$\alpha$\label{tbl-ne10shifts}}
\tablehead{
\colhead{Cross-Disp. Slice\tablenotemark{a}} & 
\colhead{Centroid Shift\tablenotemark{b}} & 
\colhead{Velocity\tablenotemark{c}}
}
 \startdata
( 20,  30) &  $2.0 \pm 1.4$  & $540 \pm 390$ \\
( 10,  20) &  $3.0 \pm 0.3$  & $800 \pm  80$ \\
(  0,  10) &  $0.4 \pm 0.5$  & $100 \pm 150$ \\
(-10,   0) &  $1.7 \pm 1.0$  & $470 \pm 280$ \\
(-20, -10) &  $1.6 \pm 0.4$  & $440 \pm 100$ \\
(-30, -20) &  $1.3 \pm 0.5$  & $350 \pm 140$ \\
 \enddata
\tablenotetext{a} {Range of cross-dispersion coordinates within the 
extracted image, in units of ACIS pixels. ACIS pixel size is 0.49 arcsec.}
\tablenotetext{b} {Shift of MEG -1 order relative to zeroth order, in units 
of ACIS pixels. Reference point (0,0) in the zero-order image was 
J2000 (RA, DEC)= (16.00885,-72.03126).}
\tablenotetext{c} {km s${}^{-1}$}
\end{deluxetable}